\documentclass{ws-procs9x6}
\usepackage{amsmath}
\usepackage{amssymb}
\usepackage{feynmf}

\newcommand {\slsh} [1] {\not{\hbox{\kern-2pt${#1}$}}}
\newcommand {\beq} {\begin{equation}}
\newcommand {\eeq} {\end{equation}}

\def\mpl{M_{\rm Pl}}
\newcommand{\gsim}{\lower.7ex\hbox{$\;\stackrel{\textstyle>}{\sim}\;$}}
\newcommand{\lsim}{\lower.7ex\hbox{$\;\stackrel{\textstyle<}{\sim}\;$}}

\newcommand {\beqn}{\begin{eqnarray}}
  \newcommand {\eeqn} {\end{eqnarray}}
\begin{document}

\title{LARGE EXTRA DIMENSIONS\,\footnote{This talk was delivered at
 UCLA
(1999), Technion (2000), DESY (2001),
University of
Pisa (2001),
UBC (2002),
NYU  (2002),
University of Nantes
(2003), Bern University (2003), Max-Planck Institute/Munich University
(2003),  
University of Padova
(2003),
University of Minnesota
(2005).}\\
Becoming acquainted with an alternative paradigm}

\author{M. SHIFMAN$^*$}
\address{THEORETICAL PHYSICS INSTITUTE\\ University of Minnesota,
116 Church Street S.E.\\
Minneapolis, MN 55455, USA\\
$^*$E-mail: shifman@umn.edu}

\begin{abstract}
This is a colloquium style pedagogical 
introduction to the paradigm
of large extra dimensions.

\end{abstract}

%\thetableofcontents

%\newpage

%\zerocounters

%\setcounter{equation}{0}
% \setcounter{subsection}{0}
% \setcounter{figure}{0}

%\newpage

%  \begin{flushright}
%{\cyr ``Ona eshche ne rodilas\cprime,}\\[0.1mm]
%{\cyr $\quad$ Ona i muzyka i slovo ..."}\\[2mm]
%{\cyr ``... Na tom rubezhe, krutom virazhe,}\\[0.1mm]
%{\cyr Na uzko{\u i} mezhe mezh eshche i uzhe."}
%  \end{flushright}

%

%%
%\renewcommand{\theequation}{19.\arabic{equation}}
%\setcounter{equation}{0}

%\renewcommand{\thesubsection}{19.\arabic{subsection}}
%\setcounter{subsection}{0}

%\renewcommand{\thefigure}{17.\arabic{figure}}
%\setcounter{figure}{0}

%

\vspace{5mm}

  \begin{center}
  \includegraphics[width=2.2in]{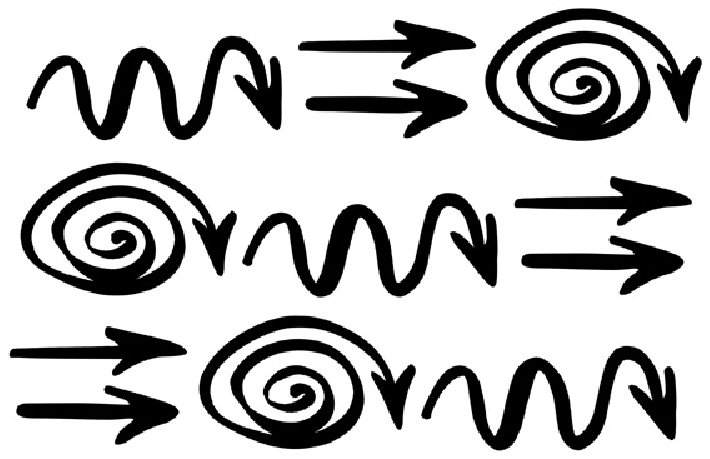}
   \end{center}

\section{Road Map (Instead of Introduction)}

They say God does not exactly know how parts of his Creation work.
When he sees a nice theory which he likes he says: ``OK, let it be so...''

Today theoretical high energy physics deals basically with two options:
(i) Grand Desert stretching from $\sim 10^2$ GeV to $\sim 10^{16}$ GeV,
with no new physics inside; and (ii) Large Extra Dimensions paradigm
various versions of which predict
new physics at a much lower scale of energies.
If the first option is realized, this would mean that high-energy physics
in the future will face a serious menace of becoming a non-empirical science:
experiments at energies in the ballpark of $\sim 10^{16}$ GeV
are impossible in terrestrial conditions.

The LED paradigm was born from the desire ``to have new physics around the corner,''
in an attempt to keep high-energy physics
as an experiment-based discipline. One may hope that God will like it.

The topic of large extra dimensions (LED) experienced an explosive
development since the mid-1990's. Since then
thousands of works dedicated to this subject were published. 
The reason why the LED paradigm attracted so much attention 
is due to the fact  that it brings the scale of new fundamental physics from
inaccessible $10^{16}$ or $10^{19}$ GeV down to 10 or 100 TeV or so.

A comparison with a huge country the exploration of which is not yet 
completed is  in order here. This lecture presents a bird's eye view of the territory,
giving a brief and nontechnical introduction to basic ideas lying behind 
the  large extra dimension paradigm and a particular braneworld model.

The task of describing the large extra dimensions paradigm is
``multidimensional'' in itself. First, there exist
three main scenarios which sometimes compete and sometimes complement each
other. Second, each scenario starts from a general design of a basic
model, while  phenomenological consequences come later. Moreover, some scenarios predict new macrophenomena, such as modifications of gravity at distances comparable
with the observed Universe size.  

%\marginpar{\frame{ \shortstack{ \rule{0mm}{1mm}\\
%{\tiny \em People hope} \\ {\tiny \em   LED to bring}  
%\\
%{\tiny \em the fundamental\,  }\\
%{\tiny \em scale of new }
%\\
%{\tiny \em physics down}\\
%{\tiny \em to $\sim$ 10 TeV.}
%\\
%\rule{0mm}{1mm}
%}}}

Below we will focus on the simplest LED scenario -- that of Arkani-Hamed--Dimopoulos--Dvali (ADD) --
limiting our forays into alternative scenarios to a minimum.
We start from a brief review of fundamental regularities of our world
in the context of the paradigm that had existed before the advent of LED.
The latter was based on the standard model and its
supersymmetric version, supersymmetric grand unification and great desert.
We will refer to this paradigm as to the great desert paradigm, or,
sometimes,  good old paradigm. (We hasten to add, though, that it was not particularly old
or particularly good.) Then, after familiarizing ourselves with the history of the
topic, we will discuss the very same regularities as they are interpreted from the standpoint of the LED
paradigm.

\section{Genesis/Glimpses of history}
\label{ggh}

\subsection{Kaluza--Klein Theory}

\begin{figure}[h]
  \begin{center}
  \includegraphics[width=3in]{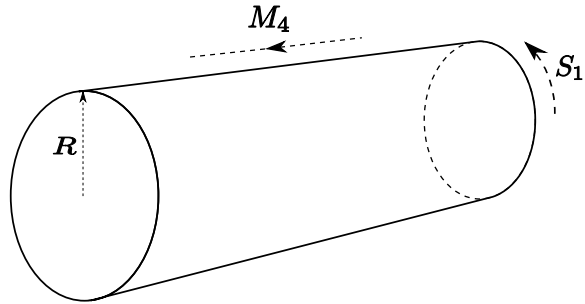}
  \caption{\small The Kaluza--Klein set-up.}
   \end{center}
\label{D1}
\end{figure}

The story starts  in the 1920's. 
At that time time Theodor Kaluza and Oscar Klein, who were working on the 
unification of Einstein's gravity
and electromagnetism,
invented the Kaluza--Klein (KK) mechanism~\cite{K1,K2}. Its essence is as follows.
Assume that our world, rather than being four-dimensional, is in fact
$(4+n)$-dimensional, $n\geq 1$, but the extra dimensions
are compact. An illustration is presented in Fig. 1, where $n=1$,
so that our world is a direct product of the four-dimensional
Minkowski space $M_4$ and a circle $S_1$ with the radius $R$.
All fields are defined on this ``cylinder.'' For a scalar field
the single-valuedness on the cylinder can be written as
\beq
\Phi (x_\mu ,\,\, Z) = \Phi (x_\mu ,\,\, Z+2\pi R)\,,\qquad \mu=0,1,2,3\,.
\label{4161}
\eeq
Here and below we will use small Latin letters for ``our'' four coordinates
reserving capital Latin letters for extra dimensions. 
The $2\pi R$ periodicity in $Z$ means that one can present the
field $\Phi (x_\mu ,\,\, Z)$ as a Fourier series,
\beq
\Phi (x_\mu ,\,\, Z) = \sum_{k=0,\pm 1,...}\, \phi_k (x_\mu ) e^{ikZ/R}\,.
\label{4162}
\eeq
The expansion coefficients depend only on ``our'' coordinates $x_\mu$,
and are often referred to as {\em modes}. As we will see shortly, the zero mode
corresponding to $k=0$ will play a special role. Modes with $k\neq 0$ 
are shown in Fig. 2. Each zero mode is accompanied by non-zero
modes $k\neq 0$ which are referred to as the KK excitations or the KK tower.

\begin{figure}[h]
  \begin{center}
  \includegraphics[width=3in]{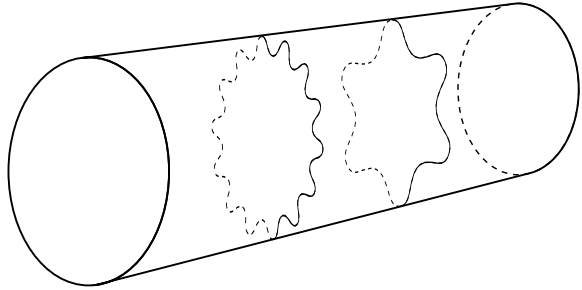}
  \caption
    {\small
The non-zero modes in the KK decomposition.
}
   \end{center}
\label{D11}
\end{figure}

From the four-dimensional point of view, the modes $\phi_k (x_\mu )$
represent a tower of regular four-dimensional fields, the so-called
Kaluza--Klein tower. Let us start from the five-dimensional wave equation, assuming that
the five-dimensional field $\Phi (x_\mu ,\,\, Z)$ is massless,
\beq
\Box_5\, \Phi (x_\mu ,\,\, Z) \equiv \left(\partial_\mu^2 -\frac{\partial^2}{\partial Z^2}
\right) \Phi (x_\mu ,\,\, Z) = 0\,.
\label{4163}
\eeq
Substituting the Fourier decomposition (\ref{4162}) we see that each mode
$\phi_k$ satisfies the four-dimensional  wave equation
\beq
\left( \Box_4\, + \frac{k^2}{R^2}\right) \phi_k (x_\mu ) \equiv \left( \partial_\mu^2 
 + \frac{k^2}{R^2}\right) \phi_k (x_\mu ) = 0\,.
\label{4164}
\eeq
The zero mode $\phi_0$ remains massless, while all other modes
become {\em massive} four-dimensional fields, with $|k|/R$ playing the role of the mass term,
\beq
m_k = \frac{|k|}{R}\,.
\label{4165}
\eeq

Compactification of $(4+n)$-dimensional fields with spin
exhibits another interesting phenomenon and leads to a richer KK tower,
which includes a spectrum of four-dimensional spins.
Consider for example the metric tensor
$G_{MN}$ in five dimensions. The vectorial indices corresponding
to higher-dimensional world here and below are denoted by capital Latin letters.
From the four-dimensional standpoint  
$G_{\mu\nu}$ is the four-dimensional metric tensor,
$G_{5\mu} = G_{\mu 5}$ is a four-vector, while $G_{55}$ is a scalar.
So, the KK tower includes the zero and non-zero modes of spin-2, spin-1 and spin-0
fields. 

In the KK picture one assumes that ``$R$ is small and $1/R$ is large''
compared to some currently available energy scale.
Moreover, all our four-dimensional world, in its entirety, including
all experimental devices and all potential observers,
is made from the zero modes. Then, given the energy limitation,
the non-zero mode quanta cannot be produced, and
 we  perceive  our world as four-dimensional.
Only when  accessible energy becomes higher  
than $1/R$ can we directly discover KK excitations, 
a signature of the extra dimension(s).

%\marginpar{\frame{ \shortstack{ \rule{0mm}{1mm}\\
%{\tiny \em Lesson \# 1:} \\ {\tiny \em   KK excitaions}  
%\\
%{\tiny \em have mass $\frac{|k|}{R}$. }\\
%{\tiny \em  Lesson \# 2:}
%\\
%{\tiny \em Only zero modes}\\
%{\tiny \em survive at $E\ll \frac{1}{R}.$}
%\\
%\rule{0mm}{1mm}
%}}}

The infancy of the Kaluza--Klein scenario
was  eventful. Suffice it to mention that
Schr\"{o}dinger, Gordon and Fock worked
on its development in the 1920's, while important
contributions in the 1930's were due to 
Pauli~\cite{pauli}, and Einstein and Bergmann~\cite{EB}.
In particular, in considering compactification of two extra dimensions
into sphere $S_2$  (see Fig. 3),  Pauli discovered the Yang--Mills theory long
before Yang and Mills. Since he did not know what to do with massless
vector fields he never published anything on this discovery.
However, gradually the interest to the Kaluza--Klein scenario languished,
probably because of the absence of realistic applications
in model-building of that time.

\begin{figure}[h]
  \begin{center}
  \includegraphics[width=3.0in]{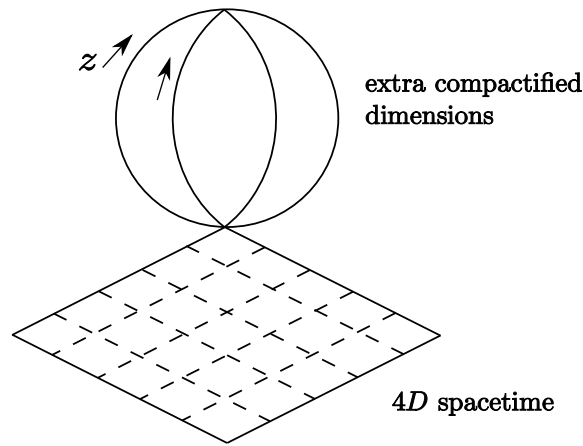}
  \caption
    {\small
Pauli considered compactification of the six-dimension space
onto $M_4\times S_2$.
}
   \end{center}
\label{pauli}
\end{figure}
 
A long period of a relative hibernation
of the Kaluza--Klein theory gave place to a revival
 in the 1980's. The dawn of a ``new era'' was marked by Witten's no-go 
theorem~\cite{wittenone}. Witten noted that  fermions cannot be chiral
if one starts from  any
eleven-dimensional manifold of the type $T=M_4\times K$
where $K$ is a compact manifold admitting the 
symmetry of the standard model (SM), namely,
SU(3)$\times$SU(2)$\times$U(1). 
This was a sad conclusion since it meant that no realistic model
could be based on the Kaluza--Klein theory 
since the fermions in our world are definitely chiral.
Fortunately, it was negated, just a few years later, 
in the first superstring ``revolution.''

\subsection{Strings}

Consistent superstring theory exists in ten dimensions.
Nonsupersymmetric string is consistent in 26
dimensions. Our world is four-dimensional.
In 1975 Joel Scherk and John Schwarz suggested \cite{JSJS} to consider
superstrings in a product space of our conventional four-dimensional space-time
and a six-dimensional compact manifold whose size is of the order of $\mpl^{-1}$.

The celebrated paper of Candelas {\em et al.}~\cite{Candelas:en},
which opened the superstring revolution of 1985,
demonstrates that if the six compact dimensions form
the so-called Calabi--Yao manifold,
then in the low-energy limit one recovers a $E(8)\times E(8)$ gauge theory which includes SM,
with three generations of chiral fermions that are observed in nature.
The Calabi--Yao compactification, conceptually,
continues Pauli's line of reasoning. As we have already mentioned,
in the 1930's Pauli observed that the KK model on $M_4\times S_2$
produces three gauge bosons of SU(2) in the low-energy
limit. The occurrence of these bosons
is due to isometries of the sphere $S_2$. Of course, the six-dimensional
Calabi--Yao manifold is much more contrived.
Geometry of the  Calabi--Yao manifold is so complicated that the explicit form 
of the metric is not known
even now. 

The typical sizes of the compact dimensions in the Calabi--Yao manifold
are of the order of $10^{-33}$ cm. Needless to say, there is no way to 
observe such extra dimensions in a direct human-designed experiment.

Systematic searches for  string-inspired realistic models of the SM type
began in the 1990's \cite{AlFa}. Since from the ``human'' standpoint,
 extra dimensions in the Calabi--Yao  scenario can be viewed as
 an auxiliary mathematical construction, 
it was suggested to replace the  compactified coordinates by 
a more formal construction ---
internal free fermions propagating on the string world sheet. One can
then completely abandon the geometrical description of the
compactification and formulate it entirely in terms of  free fermions
on the string world sheet
and their boundary conditions. One can then extract the
 physical spectrum, as well as
 the assignment of the  quantum numbers under the 
four-dimensional gauge group.

Following this procedure realistic three-generation models were
construc\-ted~\cite{AlFa}. They differed  in their detailed 
phenomenological properties,
but some elements were in common. In particular, an 
 ${\rm SO}(10)$ grand unification, with an ${\rm SO}(6)^3$ flavor
symmetries and a hidden $E_8$ gauge group were typical.

In the current millennium this topic ---
searches for  string-inspired realistic models of the SM type ---
experienced a dramatic development based on D-brane engineering \cite{ibanez}.
The advent of D branes 
\cite{jpol}
allowed one to find
string/D-brane models yielding just the SM massless fermion spectrum,
with relative ease. One of the features predicted by the D-brane-based
models  is that the SM global symmetries --- such as baryon and lepton 
numbers ---  are gauged symmetries whose anomalies are canceled (by a
     Green-Schwarz mechanism) 
only in the case of three quark-lepton generations. 
Proton stability and the Dirac nature of neutrino masses 
follow naturally. 

This direction {\em per se} --- string-inspired phenomenology ---
seems promising. 
What is important for our narrative is that
typical sizes of the compact dimensions in the string/D-brane scenario
are of the order of $10^{-33}$ cm;\,\footnote{It would be fair to add that
some work on introducing larger extra dimensions in the string context
had been done in the 1990's. 
For instance, in 1990 Antoniadis suggested \cite{Anton} 
large extra dimensions in the context of the
Standard Model, with gauge fields 
propagating in the bulk and matter fields
localized on the orbifold fixed points (although the word brane was not used).
Somewhat later, Ho\v{r}ava and Witten 
pointed out \cite{horwi} that a single extra dimension
of the size $\sim 10^{-28}$ cm could eliminate the gap between the scale of 
grand unification and the Planck scale, see Sect. \ref{hieros}.}
 hence, the masses of the excited states
in the KK tower are of order of $10^{19}$ GeV.  Such energies are (and will be) 
inaccessible to any terrestrial experiment. In other words, if physics were to be described
by such scenarios, the KK tower would be  unobservable,
and the KK theory would have no practical implications.

\subsection{Parallel development: localizing on topological defects}

Independent ideas which later formed one of the pillars of the
LED paradigm emerged and were developed in the 1980's-90's. The idea of 
localizing matter on
topological defects was formulated, in the most clear-cut form,\footnote{Similar ideas 
were discussed around the same time in Refs. 
[13,14].}
%~\cite{Akama,Visser}.} 
by Rubakov and Shaposhnikov
\cite{Rubakov:bb} in the paper entitled ``Do We Live Inside A Domain Wall?''.

It is convenient to explain the essence of this suggestion in a simplified setting
where ``our'' world is assumed to be (1+2)-dimensional, while the
coordinate $z$ is treated as an ``extra dimension.'' Assume that the underlying
microscopic theory has several discrete degenerate vacua which are labeled 
by distinct values of an order parameter. Call two such vacua --- they can
be chosen arbitrarily ---  vacuum I and II. There exists a static field configuration,
a domain wall, which divides the three-dimensional space in two parts,
say, on the left hand-side our system is in the vacuum I while 
on the right-hand side in the vacuum II (Fig. 4). 

\begin{figure}[h]
  \begin{center}
  \includegraphics[width=2.5in]{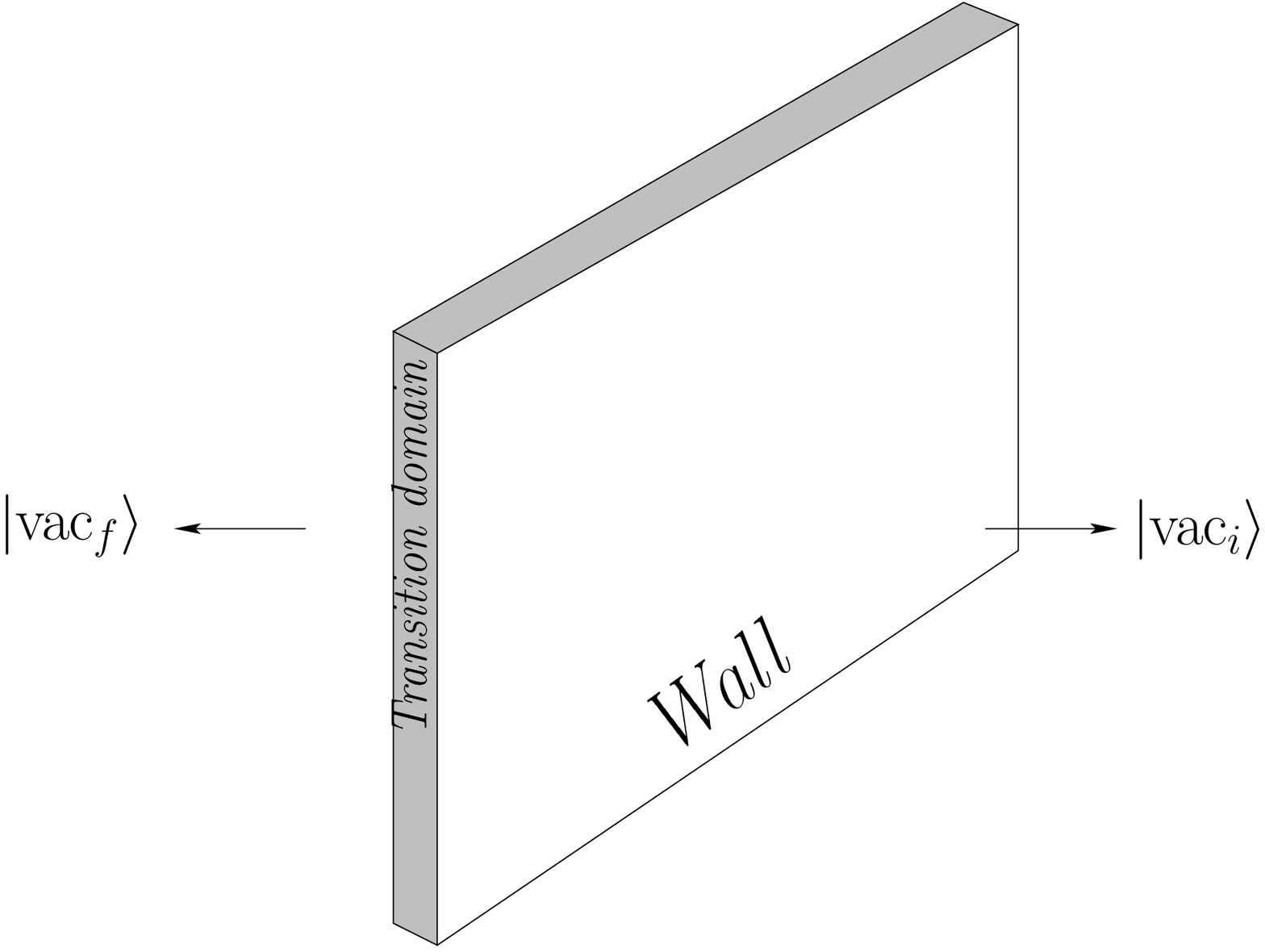}
\caption{A domain wall separating two distinct degenerate vacua.}
   \end{center}
\label{dwa}
\end{figure}

The domain wall represents
a transitional domain and is topologically stable.
Once created, it cannot be destroyed. The thickness of the domain wall
$\delta$ depends on details of the microscopic theory.
At distances $\gg \delta$ the domain wall can be viewed as a two-dimensional surface.

One can excite the domain wall field configuration
by pushing the wall at a certain point or by pumping in energy
in any other way. All possible excitations naturally fall into two
categories. Some of them are localized on the wall (their spatial extension in the
perpendicular direction is of the order of $\delta$). These are usually associated 
with {\em  zero modes}. Being considered from the (1+2)-dimensional point
of view, the zero modes represent massless particles which can propagate only along
the wall surface. 

Other excitations are delocalized and can escape in the bulk (i.e. in the perpendicular
direction). They are represented by
{\em nonzero} modes with typical energy
eigenvalues of the order of $1/\delta$. From the (1+2)-dimensional standpoint
each nonzero mode is a particle with mass $M_n \sim 1/\delta$.

Assume that all matter that we see around 
is made of the zero modes trapped on the domain wall surface.
Then ``our world'' will be confined to the wall surface and will be effectively
(1+2)-dimensional. To discover the third (perpendicular) spatial dimension an
observer made of the zero modes will have to have 
access to energies larger than $1/\delta$. 

An obvious distinction between the KK scenario and localization on the domain walls
(or other topological defects) is the mass scale of the excited modes.
In the KK model it is related to the inverse size of the extra dimension,
while in the case of the domain walls the extra dimension is infinite,
and the mass scale is set by the inverse thickness of the wall. 
This distinction turns out to be  crucial in physical applications.

%\marginpar{\frame{ \shortstack{ \rule{0mm}{1mm}\\
%{\tiny \em Lesson \# 3:} \\ {\tiny \em   Zero modes}  
%\\
%{\tiny \em trapped on a wall }\\
%{\tiny \em are separated from}
%\\
%{\tiny \em nonzero modes}\\
%{\tiny \em by  $E \sim \frac{1}{\delta}.$}
%\\
%\rule{0mm}{1mm}
%}}}

The existence of at least one zero mode is easy to demonstrate.
Indeed, the underlying microscopic theory has four-dimensional
translational invariance. The domain wall breaks, spontaneously, the invariance
with respect to translations in the $z$ direction. Physics becomes dependent
on the distance to the wall in the perpendicular direction.
Correspondingly, in accordance with the Goldstone theorem,
there emerges a Goldstone boson which is confined to the wall surface.
If the profile of the order parameter describing the wall (we will call it $\phi (z)$)
is known, then the profile of the translational zero mode is given by the derivative
$d \phi /dz$, see Fig.~5.

\begin{figure}[h]
  \begin{center}
  \includegraphics[width=3.0in]{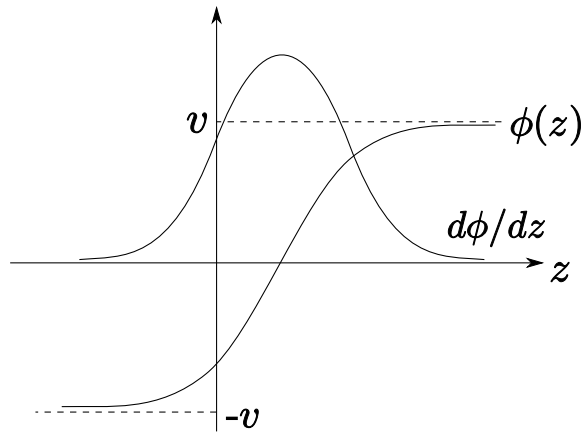}
\caption{The profile function of the domain wall determines localization
of the translational zero mode.}
   \end{center}
\label{dwap}
\end{figure}
 
The (1+2)-dimensional Goldstone boson appearing in this way has spin zero.
In fact, the original work of Rubakov and Shaposhnikov was
motivated by the desire to have a Higgs boson whose mass is protected
by the Goldstone theorem from being dragged in the ultraviolet by
 quadratic divergences, typical of the scalar particle masses in field theory.
The novelty of the idea and its potential were not recognized till mid-1990's
since shortly after the Rubakov--Shaposhnikov publication
supersymmetry gained the role of a universal saviour.

In general, localization of spin-zero bosons presents no problem, at least, at 
a conceptual level. Assume that the microscopic theory has a global symmetry 
group $G$ which remains unbroken both in the vacuum I and II.
Assume that on the given wall solution the symmetry $G$ is (spontaneously)
broken down to $H$. Then the Goldstone bosons corresponding to
the broken generators will be confined to the wall;
their interactions will be described by a coset $G/H$ sigma model.
%
%\marginpar{\frame{ \shortstack{ \rule{0mm}{1mm}\\
%{\tiny \em The number of} \\ {\tiny \em   the Goldstones}  
%\\
%{\tiny \em is  }
%\\
%{\tiny \em \,  {\rm dim}$_G$ - {\rm dim}$_H$. }
%\\
%\rule{0mm}{1mm}
%}}}

Localization of the spin-1/2 particles
is also a long-known phenomenon.
Fermion fields coupled to the wall can have
zero modes too. The number of such zero modes is regulated
by the Jackiw--Rebbi index theorem \cite{JR}.
For fermion fields one must consider the mass matrix as a function of
$z$. If in piercing the wall
(i.e. in passing from $z=-\infty$ to $z=\infty$)
$k$ eigenvalues of the fermion mass matrix change sign,
then one will have $k$ fermion zero modes. The thickness of the profile of the 
fermion zero modes is of the order of the inverse fermion mass in the bulk.
With all bulk fermions massive, localization of the fermion zero modes on the wall clearly presents no problems.

Localizing non-Abelian gauge fields on the wall (in the framework of
field theory) is a more complicated task. A working mechanism was found
in 1996 \cite{Dvali:1996xe}. The basic idea is as follows.
Assume that in the bulk, outside the wall, we have a gauge theory (with the gauge group $G$)
in the confining phase. The bulk dynamical scale parameter is $\Lambda_{\rm bulk}$.
Assume that inside the wall the gauge group is $G^\prime$ where $G^\prime\in G$,
the dynamical scale is $\Lambda^\prime$, and $\Lambda^\prime\ll \Lambda_{\rm bulk}$.
Then the gauge fields of the $G^\prime$ theory will be localized on the wall.
Indeed, at energies $\Lambda^\prime\ll E\ll\Lambda_{\rm bulk}$
 the chromoelectric  fields of the $G^\prime$ theory cannot escape in the bulk since,
 due to confinement, the lightest states in the bulk (glueballs)
 have masses $\sim \Lambda_{\rm bulk}$.
 One can dualize this picture. Assume that the bulk $G$ theory is Higgsed.
 Then the probe magnetic charges are confined in the bulk through formation of chromomagnetic flux tubes. If appropriate condensates vanish inside the wall, then the chromomagnetic flux 
 can spread freely inside the wall.

In string theory localization of the gauge fields is achieved
with no special effort, provided one identifies
domain walls with $D$ branes. The gauge bosons are then represented by open strings
with the end points attached to the branes (Fig. 6). Thus, they are naturally
confined to the brane surface.
(Let us parenthetically note that the closed strings representing gravitons can freely propagate in the bulk in this picture. We will return later to the discussion of this feature.)

\begin{figure}[h]
  \begin{center}
  \includegraphics[width=2.5in]{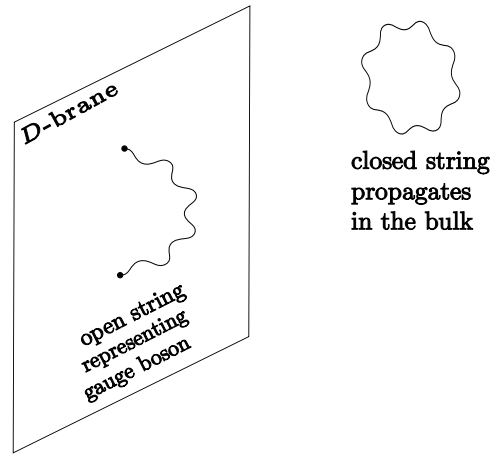}
\caption{The gauge bosons are represented by open strings and are attached to the $D$-brane,
while the gravitons represented by the closed strings ``live" in the bulk.}
   \end{center}
\label{closed}
\end{figure}

As was mentioned, the idea of localization on topological
defects was in a rather dormant state till mid-1990's. 
In 1996 it was revived in the supersymmetric
context \cite{Dvali:1996bg}. 

Why supersymmetry? 

The reason for invoking 
supersymmetry is two-fold. First, in nonsupersymmetric theories
the existence of degenerate discrete vacua requires
 spontaneous breaking of some discrete symmetry.
On the other hand, in supersymmetric theories
the vacuum degeneracy is hard to avoid, it is typical.
Indeed, all supersymmetric vacua must have the vanishing energy density
and are thus degenerate. Therefore, domain walls are more abundant.

The second motivation
 is the ease with which one can localize simultaneously
 spin-0 and spin-1/2 fields on critical (or BPS saturated) domain walls.
%of a special type.
%Usually they are referred to as Bogomolny--Prasad--Sommerfield-saturated \cite{BPS}, 
%or BPS for short.\,\footnote{Sometimes the BPS objects are called critical.
%This is especially common in the literature dealing with gravity
%or supergravity.} 
%What does that mean?

%The anticommutation relation defining supersymmetry in four dimensions
%has the form~\cite{GL}
%\beq
%\{\bar Q_{\dot\alpha}\,,\, Q_\alpha\} = 2 P_{\dot\alpha \alpha}\,,\qquad \alpha ,\dot\alpha
%=1,2\,,
%\label{gel}
%\eeq
%where $\bar Q_{\dot\alpha}$ and $Q_\alpha$ are the supercharges while
%$P_{\dot\alpha \alpha}$ is the energy-momentum operator, which is the 
%generator of  translations in space and time.
%Since the domain wall breaks the translational invariance
%in the $z$ direction, typically  all four supercharges are broken,
%and the effective low-energy theory on the wall surface
%turns out to be non-supersymmetric. However, in special cases
%the defining equation (\ref{gel}) gets modified (a {\em central charge}
%appears on the right-hand side), and then one can protect, say, two out of four
%supercharges from breaking. Then each boson zero mode must be 
%accompanied by a fermion one,
%and {\em vice versa}. The theory of the states localized 
%on the wall is then  supersymmetric.

\subsection{What to do with gravity?}
\label{wtdwg}

One cannot hope to successfully describe our world without
including gravity. Unlike spin-0,1/2 and 1 fields,
no field-theoretic mechanisms ensuring {\em bona fide} localization
of  gravity on
domain walls are known.
Moreover, if one approaches the problem from the string theory
rather than field theory side, a drastic distinction
between, say,  gauge fields and gravity is obvious too.
The gauge fields are represented by open strings with the endpoints
attached to $D$ branes. Thus, they are naturally localized on
$D$ branes. At the same time, gravity is represented
by closed strings which can freely propagate in the bulk,
see Fig. 6.
In order for the LED paradigm to take off, a fresh idea regarding what
to do with gravity was needed. It was not before long that it was put forward.
%
%\marginpar{\frame{ \shortstack{ \rule{0mm}{1mm}\\
%{\tiny \em Let gravity} \\ 
%{\tiny \em   go in the }  
%\\
%{\tiny \em bulk }
%\\
%\rule{0mm}{1mm}
%}}}

\section{Flat compact extra dimensions}
\label{ced}

Historically the first was the compact extra dimension
model which goes under the name ADD, where ADD stands for
Arkani-Hamed, Dimopoulos and Dvali, its inventors
\cite{ADD1}. 
In the March 1998 paper, and a follow-up publication
of the same authors with Antoniadis
\cite{ADD2},  a marriage between the Kaluza--Klein scenario and localization
on domain walls was suggested. 
 Compactifying extra dimensions a l\'{a}
 Kaluza--Klein solves the problem of gravity, while localizing
all other fields on the wall solves the hierarchy problem.

But let us begin from the very beginning.

\subsection{The Arkani-Hamed--Dimopoulos--Dvali (ADD) scenario}
\label{addsc}

The model starts from the assumption that the space-time is $(4+n)$-dimensio\-nal,
$n\geq 1$, while its geometry is factorized,
\beq
M_{\rm world} = M_4\times K_n\,.
\label{factg}
\eeq
All SM particles are localized on a $(1+3)$-dimensional domain wall
(3-brane) representing  $ M_4$ in the above expression
(Fig. 7). At the same time,
 gravity spreads to all $4+n$ dimensions
 (Fig. 8).
  \begin{figure}[h]
  \begin{center}
  \includegraphics[width=3in]{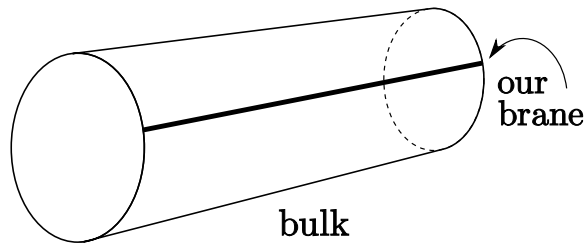}
\caption{``Our world" is the 1+3 dimensional domain wall which is shown by a solid line in this figure.
Perpendicular directions are compact.}
   \end{center}
\label{closed2}
\end{figure}
 \begin{figure}[h]
  \begin{center}
  \includegraphics[width=3in]{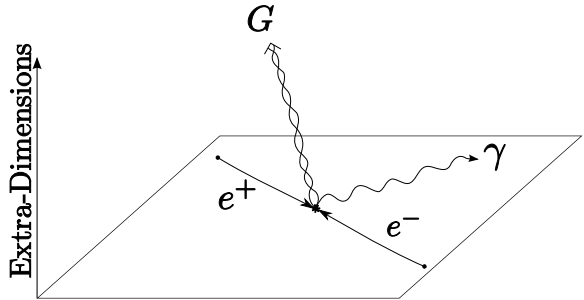}
\caption{We, and everything around us, are made from the zero modes localized on the
domain wall. Gravity escapes in the bulk.}
   \end{center}
\label{closed3}
\end{figure}
 
If the thickness of the
domain wall is chosen to be $\delta \lsim (10\,\,  {\rm TeV})^{-1}$
then at energies $ \lsim  10 \, \, {\rm TeV}$ physics is effectively four-dimensional
in all experiments except those with gravity.\footnote{One may ask where does this mass scale, 
10 TeV, come from? We will answer this question shortly.}
Since gravity escapes in the bulk it becomes four-dimensional only at distances
$r\gg R$. At distances $r\lsim R$ gravity is $4+n$ dimensional.

We want $R\gg \delta$. The reason behind this requirement will become clear
momentarily. Let us ask ourselves what values of the size of the compact
dimensions $R$
are compatible with what we know about our world today.
Constraints on $R$ are surprisingly lax. They follow from the fact that in the ADD scenario
gravity becomes effectively four-dimensional at distances
$\gsim R$. Experimentally, gravity is well-studied at large distances
(where it is certainly four-dimensional)
and known to a much lesser extent at short distances.
In fact, below 0.1 mm or so the gravitational force has not been measured, and one cannot rule
out that at such distances it is  $(4+n)$-dimensional.\,\footnote{In 1998 
measurements of the gravity  extended down to 1 mm. Dedicated experiments
\cite{Hoyle:2000cv}
performed after the ADD suggestion gave rise to a considerable theoretical activity
improved the above bound by an order of magnitude.}

Let us assume that the size of extra dimensions $R\sim 0.1$ mm and see
to what consequences this assumption leads. If in the future gravity will be proved 
to be four-dimensional at such distances one can always downsize extra dimensions
making  $R \sim 0.01$ mm or less. 

\subsection{Fundamental scale and the size of extra dimensions}
\label{fssed}

Somewhat symbolically, the action can
be written as
\beq
S=\frac{M_{\rm f}^{2+n}}{2}\,\int d^4x \int_0^{2\pi R}d^n Z\, 
\sqrt{G}\, R_{4+n} + \int d^4x \sqrt{g}\left( T + {\cal L}_{\rm SM}(\Phi_{\rm SM})\right)
\label{SADD}
\eeq
where $M_{\rm f}$ is the {\em bona fide} fundamental
constant of gravity, $G$ and $R_{4+n}$ are the $(4+n)$-dimensional
metric and scalar curvature, respectively,
$g$ is four-dimensional metric, and, finally,
${\cal L}_{\rm SM} $ is a Lagrangian describing
all SM fields (which, remember, are trapped on the brane).
A typical mass scale associated with ${\cal L}_{\rm SM}$ will
be denoted as $M_{\rm SM}$,
$$
M_{\rm SM} \sim 100\,\,{\rm GeV}\,.
$$
The constant $T$ has to be adjusted in such a way
that the overall cosmological term, which includes $T$ plus all quantum loops,
vanishes. This is the usual fine-tuning condition on the cosmological constant.
The ADD scenario adds nothing new in this respect.

Now let us apply the Kaluza--Klein mode expansion to the graviton field
and keep, for the time being, only the zero mode, neglecting all states in the KK tower
with masses of the order of  $1/R$. Since the zero mode is $Z$ independent,
we can perform the $Z$ integration in the first term on the right-hand side
thus obtaining
\beq
{M_{\rm f}^{2+n}\over 2}\,\int d^4x \int_0^{2\pi R}d^n Z\, 
\sqrt{G}\, R_{4+n}
\longrightarrow 
\frac{1}{2}\, M_{\rm f}^{2+n}\, V_n
\int d^4x \, \sqrt{g}\, R \,,
\label{zmgravity}
\eeq
where $g$ and $R$ are four-dimensional metric and scalar curvature evaluated on the
zero mode, and $V_n$ is the volume of extra dimensions,
\beq
V_n = \left( 2\pi R \right)^n\,.
\eeq 
The expression on the right-hand side of Eq.~(\ref{zmgravity})
is applicable at distances $r\gg R$.

At such distances  the gravitational potential
takes the standard Newton form
\beq
V(r)=-{ G_N m_1m_2 \over r} \,,
\label{Vrlarge}
\eeq
with the Newton constant$G_N$ determined by Eq.~(\ref{zmgravity}).
\beq
 G_N=(M_{\rm f}^{2+n}\, V_n)^{-1}\,.
 \label{e111}
 \eeq
At the same time, if $r\ll R$ we 
must return to Eq.~(\ref{SADD}) which implies the following static
potential
\beq
V(r)=-{ m_1m_2 \over M^{2+n}_{\rm f}\, r^{1+n}} \,.
\label{Vshort}
\eeq
It is quite obvious that, upon inspecting
Eq. (\ref{e111}),  a four-dimensional observer
will interpret $ M_{\rm f}^{2+n}\, V_n$ as the Planck scale,
\beq
M_{\rm f}^{2+n}\, V_n = M_{\rm Pl}^2 \sim \left( 10^{19}\,\, {\rm GeV}\right)^2\,.
\label{exdvol}
\eeq
He or she will think that this is the fundamental scale at which gravity becomes 
of order unity and needs quantization. The ``visible" fundamental scale, as established
by such an observer, is
separated from  accessible energies $M_{\rm SM}$
by a huge interval, thus creating an enormous hierarchy of scales.

In fact, in the ADD scenario the genuine fundamental scale is $ M_{\rm f}$;
this is the energy at which $4+n$ dimensional gravity becomes strong. 
$ M_{\rm f}$ is related
to the ``visible" fundamental scale $M_{\rm Pl}$ as follows:
\beq
M_{\rm f}= \frac{1}{2\pi R }\,\left(2\pi R\,  M_{\rm Pl} \right)^{\frac{2}{n+2}}
\eeq
If $n=1$ and $M_{\rm f}\sim$ 10 TeV, then 
$R\lsim 10^{12}$ cm which is definitely inconsistent
with the Newton law well-established at such distances. 
Thus,  a  single extra dimension is ruled out in the ADD scenario. Then it is natural to assume
that $n=2$. If so, and $M_{\rm f}\sim$ 10 TeV, we can use the formula
\beq
R = \frac{1}{2\pi \, M_{\rm f} }\left(\frac{M_{\rm Pl}}{M_{\rm f}}
\right)^{2/n}
\eeq
to deduce that $R\sim$ 0.1 mm.

Thus, two or more extra dimensions in the ADD framework 
are not inconsistent with the existing data. 
What is important is that even at larger $n$ the size of the extra dimensions
 $R$ is  large compared with $\delta^{-1}$ and the more so with $M_{\rm Pl}^{-1}$.
There are no theoretical motives behind the choice
of $M_{\rm f}\sim$ 10 TeV. The only reason is the desire to have new physics in the accessible
range of energies. If a typical scale of new physics is higher than 10 TeV,
one can adjust $M_{\rm f}$ appropriately.

Several simplifying assumptions were made in the course of the above 
consideration, namely,

(i) The wall thickness $\delta$ (which is assumed to be $\delta \sim M_{\rm f}^{-1}$)
is neglected;

(ii) The wall shape fluctuations are neglected (these are the Nambu--Goldstone bosons
which couple to matter derivatively);

(iii) All extra dimensions are assumed to have equal size $R$;

(iv) It is postulated that only gravity propagates in the bulk.

One or more of the above assumptions
can be easily lifted. For example, one can consider several 
extra dimensions with individual sizes, or let escape in the
bulk other fields in addition to gravity (but only those which 
carry no charges with respect to SM). This will change technical 
details, leaving conceptual foundations of the approach intact.

\subsection{Phenomenological implications}
\label{phenim}

Some basic 
phenomenological regularities that we observe in Nature
are:

\vspace{1mm}

\indent
(i) Proton stability;\\
\indent
(ii) Mass hierarchies;\\
\indent
(iii) The existence of three and only three generations;\\
\indent
(iv) The lightness of neutrinos;\\
\indent
(v)  The lightness of (yet to be discovered)  Higgs particles;\\
\indent
(vi) A peculiar pattern of
 quark mixing angles (the CKM matrix \\
$~~~~~~~~~~~$elements); a peculiar pattern of the neutrino mixing angles.

\vspace{1mm}

In the years that elapsed after the creation of the standard
model and before the advent of the LED paradigm these  regularities received
more or less satisfactory explanations.
Some of them were understood  at a conceptual  level,
while for others detailed technical explanations were 
worked out.  At the very least, we believed that the above regularities
presented  no  mysteries that could shake  foundations of physics.
For instance, the proton stability was explained by a very high mass scale
of unification/strong gravity, of the order of $M_{\rm Pl}$.
The lightness of the left-handed neutrino was thought to be due to a {\em  seesaw}
mechanism (see below), and so on.

With the advent of the LED paradigm a drastic rethinking of
particle physics, and in particular, flavordynamics, was inevitable.
Everything had to be questioned anew, and novel explanations
had to be invented. Needless to say, they were suggested before long.

Below we will outline some mechanisms discussed in this context.
 The aspect which we will emphasize
here is a natural conversion of dynamical regularities into
geometric ones within the LED framework.

\subsection{Hierarchy of scales}
\label{hieros}

At terrestrial energies gravity is exceedingly weak.
The gravitational interaction is characterized by the Newton constant
$G_N\equiv M_{\rm Pl}^{-2}$. While a typical electroweak scale is $M_{\rm SM}\sim 100$
GeV, the Planck scale $ M_{\rm Pl} \sim 10^{19}$ GeV, so that there
is a huge hierarchy of scales, $M_{\rm Pl} / M_{\rm SM} \sim 10^{17}$.

In the standard model {\em per se}, the electroweak scale is not stable,
since quantum loops drag the Higgs boson mass (which is supposed to be of the order of
$M_{\rm SM}$) to the Planck scale. In the desert paradigm
supersymmetry plays a protective stabilizing role ---
superpartners  cancel quadratic divergences in the
Higgs boson mass above the supersymmetry breaking scale $M_{\rm SUSY}$.
Although supersymmetry is not yet discovered experimentally,
the general belief is that $M_{\rm SUSY}\sim $ few hundred GeV, i.e. quite
close to $M_{\rm SM}$.

  \begin{figure}[h]
  \begin{center}
  \includegraphics[width=2.5in]{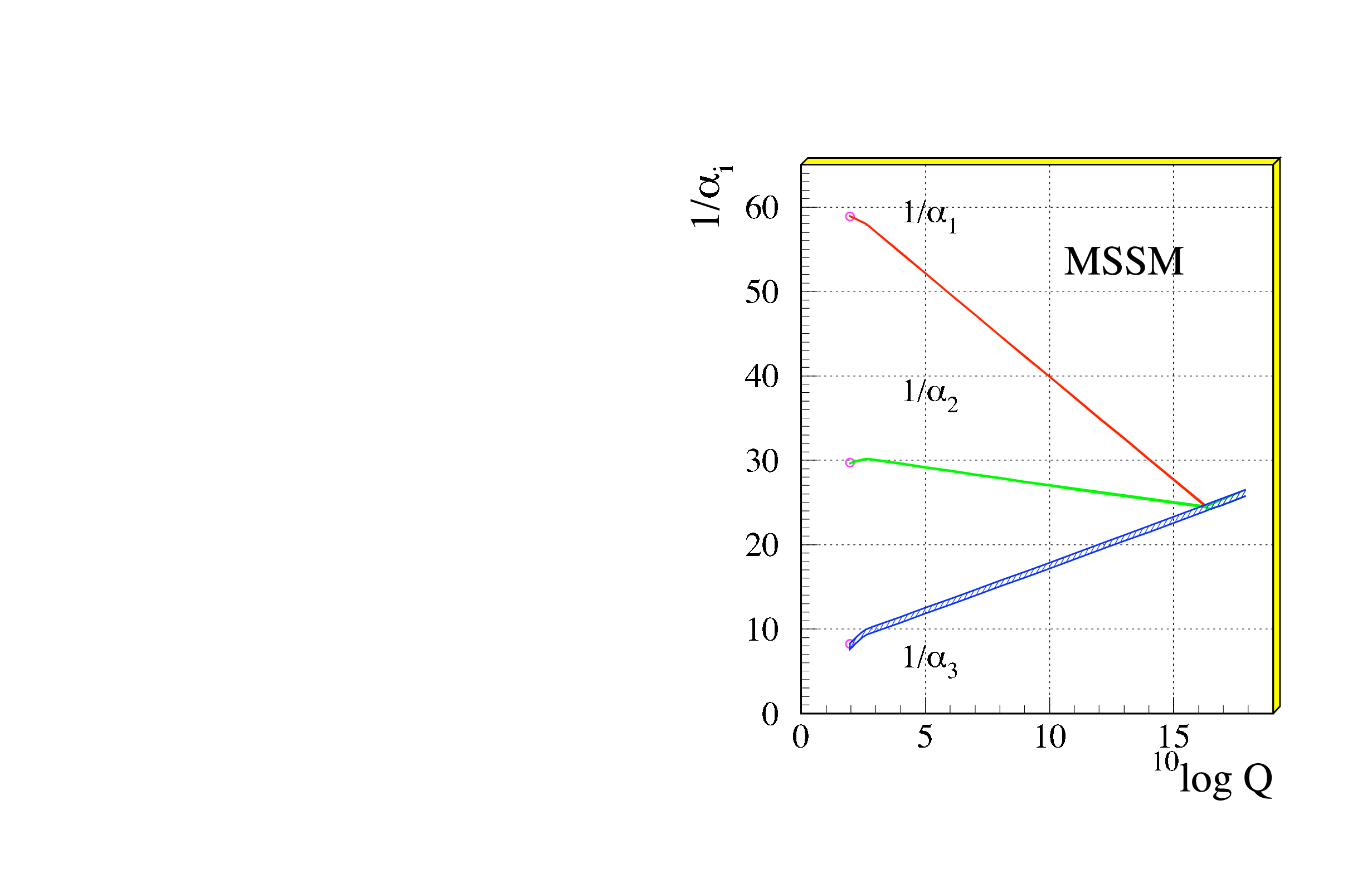}
\caption{Evolution of the inverse of the three gauge coupling constants
in  MSSM.}
   \end{center}
\label{closedp}
\end{figure}

A minimally  supersymmetrized version of SM (minimal supersymmetric standard model, 
or MSSM for short) has three gauge groups,  U(1), SU(2), and SU(3), and,
correspondingly, three gauge coupling constants, $\alpha_1$, $\alpha_2$,
and $\alpha_3$. These coupling constants run; the running formula
being  logarithmic in energy. 
 This slow logarithmic running introduces 
another huge scale --- the scale of grand unification. In MSSM the scale of grand unification
$M_{\rm GUT}\sim 10^{16}$ GeV. It turns out
that  all three gauge coupling constants unify to within
a few percent at $M_{\rm GUT}$, see Fig. 9.
The grand unification scale is, in turn, quite close to the Planck scale $M_{\rm Pl}$,
where gravity becomes strong,
\beq
M_{\rm GUT} \sim 10^{-3} M_{\rm Pl}\,.
\label{mgut}
\eeq
There is a vast desert between $M_{\rm SM}$ and $M_{\rm SUSY}$ on the one hand,
and $M_{\rm GUT}$ and $M_{\rm Pl}$ on the other hand, stretching in energy over
16 to 17 orders of magnitudes.
%
%\marginpar{\frame{\shortstack{ 
%\rule{0mm}{1mm}\\
%{\tiny \em ``... looking up at} 
%\\ 
%{\tiny \em   the Planck mass of}  
%\\
%{\tiny \em $10^{19}$ GeV, we are in  }
%\\
%{\tiny \em a position of greater}
%\\
%{\tiny \em inferiority than an }
%\\
%{\tiny \em ant staring up at a}
%\\
%{\tiny \em skyscraper (facing }
%\\
%{\tiny \em a factor of only}
%\\
%{\tiny \em $10^6$ or so)...'' }
%\\
%{\tiny \em  M. Gell-Mann}
%\\
%\rule{0mm}{1mm}
%}}}

In the LED scenario there is only one fundamental scale,
$M_{\rm f}\sim 10$ TeV. That's the scale where gravity must become strong,
and all interactions unify.\footnote{Beyond $M_{\rm f}$ physics is $4+n$ dimensional, and the law
of running of the gauge couplings is non-logarithmic \cite{gher}. }
 Where has the hierarchy gone?

The enormous hierarchy did not disappear. The energy hierarchy
 is converted into
a geometric hierarchy of the ``transverse'' sizes: the radius of the extra dimension 
versus the brane thickness.  Why the compact dimensions are so large in the
LED paradigm? It is impossible to answer this question without complete
understanding of the compactification dynamics, which is well beyond the scope of this lecture.

\subsection{Proton stability}
\label{prost}

Why this issue is potentially dangerous for the very existence of the ADD
scenario?

To properly set a reference point, let us start
from the desert paradigm, where two mechanisms were considered 
in connection with the problem of the proton decay. 
First, in the grand unification theories (GUT's), there exist
gauge bosons
with mass $\sim M_{\rm GUT}$ and  leptoquark 
quantum numbers (in the Russian literature they are sometimes called ``elephants'').
These elephant  bosons mediate  quark annihilation into leptons,
see Fig. 10,
 leading to the proton decay ($B-L$ is still conserved).
Experimentally the proton lifetime is known to exceed
$10^{32}$ years. The appropriate suppression
of the ``elephant'' contribution is due to the fact that their
 masses are very large,
  $\gsim M_{\rm GUT}$.

\vskip 2mm

\begin{figure}[h]
\begin{center}
\begin{fmffile}{feyn_ch10_1}
\begin{fmfgraph*}(200,120)
   \fmfleft{i1,i2}\fmfright{o1,o2}
  \fmf{fermion}{i1,v1,i2}
  \fmf{fermion}{o1,v2,o2}
  \fmf{photon,label=$\chi$}{v1,v2}
  \fmfdot{v1,v2}
  \fmflabel{$q$}{i1}
  \fmflabel{$e^c$}{i2}
  \fmflabel{$q$}{o1}
  \fmflabel{$u^c$}{o2}
\end{fmfgraph*}
\end{fmffile}
\end{center}

\vskip 7mm

\begin{center}
\begin{fmffile}{feyn_ch10_2}
\begin{fmfgraph*}(200,120)
   \fmfleft{i1,i2}\fmfright{o1,o2}
  \fmf{fermion}{i1,v1,i2}
  \fmf{fermion}{o1,v2,o2}
  \fmf{photon,label=$\chi$}{v1,v2}
  \fmfdot{v1,v2}
  \fmflabel{$\ell$}{i1}
  \fmflabel{$d^c$}{i2}
  \fmflabel{$q$}{o1}
  \fmflabel{$u^c$}{o2}
\end{fmfgraph*}
\end{fmffile}
\end{center}

\caption{Diagrams responsible for the proton decay of the type $p \to \pi^0 e^+$, or 
$p \to K^0 e^+$, or $p \to K^+ \nu$ in supersymmetric GUT's. ``Elephants" are denoted by $\chi$.}

\label{protdec}
\end{figure}
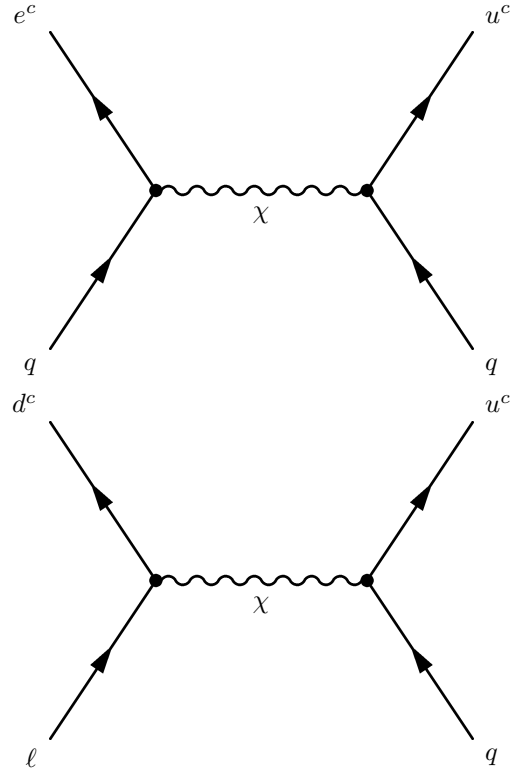

The proton decay rate associated with this mechanism (for a review see e.g. [23,24].)
is easy to estimate,\,\footnote{The estimate given below
refers to the so-called dimension-six operators, see Fig. 10; dimension-five
operators result in a different (more contrived) expression,
which is close numerically, however.}
\beq
\Gamma_{\rm proton} \sim \alpha^2 \, m_{\rm proton}\, 
\left( \frac{m_{\rm proton}}{M_{\rm GUT}}\right)^4\,,
\label{protlt}
\eeq
where $\alpha$ is the common value of the three gauge couplings
at the unification scale. 
Given Eq. (\ref{mgut}), the proton lifetime is predicted to be
longer than $10^{33}$ years, which is compatible with experiment.

 \begin{figure}[h]
  \begin{center}
  \includegraphics[width=3in]{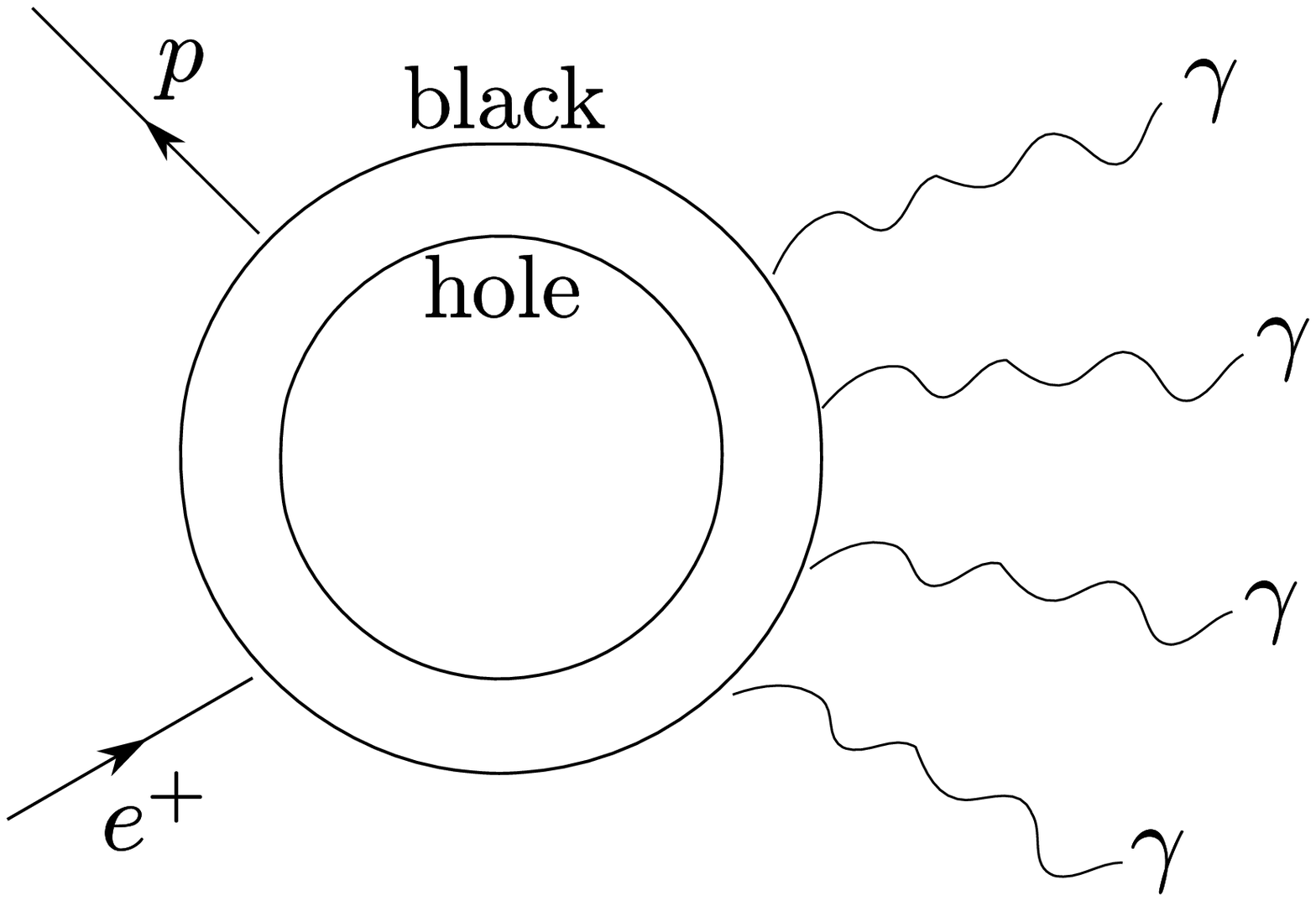}
\caption{Proton decay through a virtual black hole.}
   \end{center}
\label{prohole}
\end{figure}
 \begin{figure}[h]
  \begin{center}
  \includegraphics[width=3in]{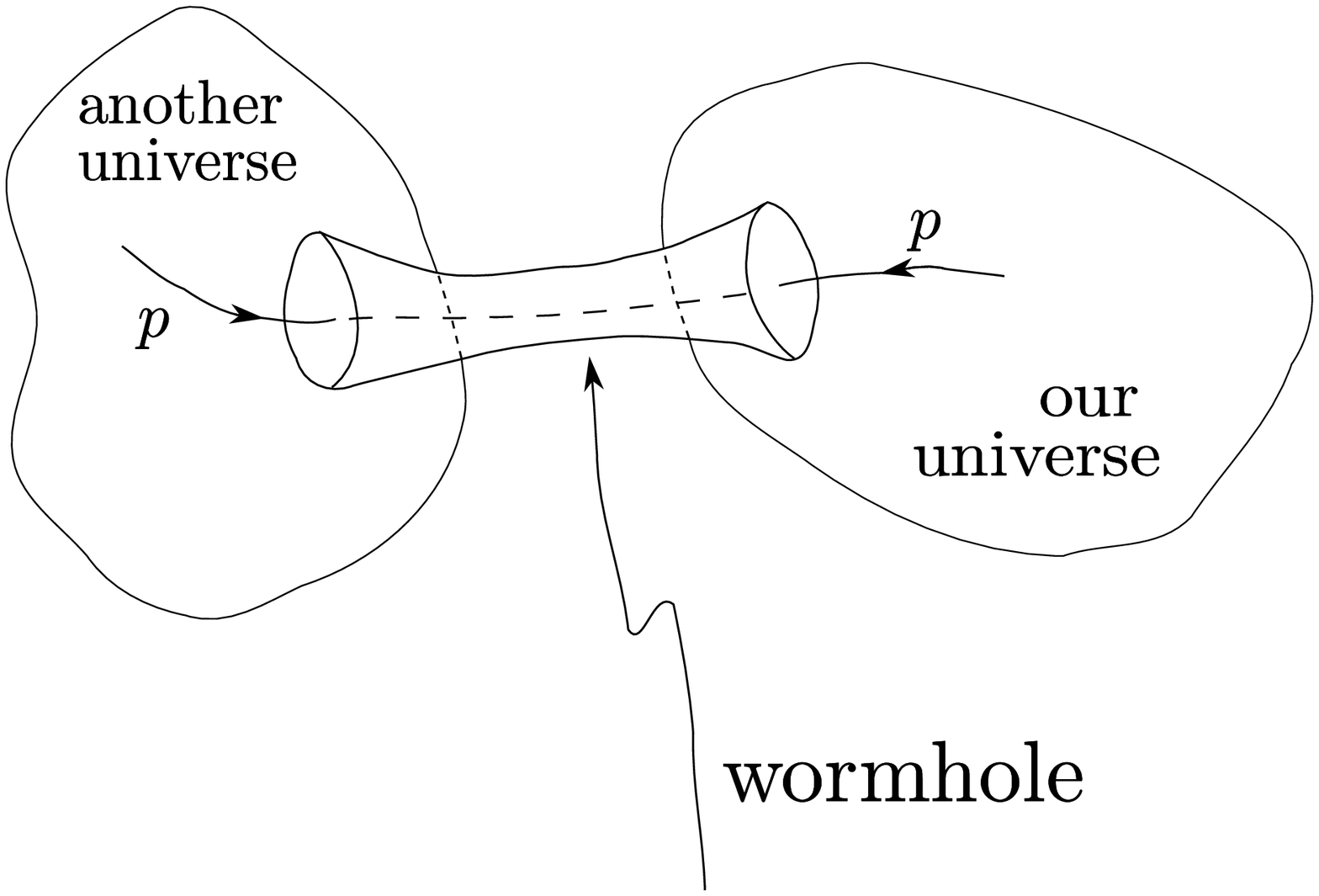}
\caption{A proton escaping from our universe via a wormhole.}
   \end{center}
\label{proholep}
\end{figure}

The second mechanism discussed in connection with the
proton decay is possible non-conservation of global quantum numbers,
such as the baryon number $B$ or lepton number $L$,
in the presence of quantum gravity.
The argumentation goes as follows  \cite{blholep}:
global charges can be swallowed by black holes ---  say, virtual 
black holes which are formed non-perturbatively at the distance scales
where gravity becomes strong --- which then eventually evaporate (Fig. 11). 
Or a wormhole may suck in a global charge from our universe (Fig. 12)
and spit it out into another one \cite{wormhole}.
In both cases such event will be interpreted by ``our'' observer 
as a proton decay.
In the ``old'' four-dimensional picture the
contribution of
this mechanism is of the order of
\beq
\Gamma_{\rm proton} \sim  \, m_{\rm proton}\, 
\left( \frac{m_{\rm proton}}{M_{\rm Pl}}\right)^4\,,
\label{protltp}
\eeq
a rate which is even smaller than that in Eq. (\ref{protlt}).
What makes the proton instability due to gravity effects
phenomenologically acceptable is the huge value
of the mass scale where gravity becomes strong.
%
%\marginpar{\frame{\shortstack{ 
%\rule{0mm}{1mm}\\
%{\tiny \em Black holes} 
%\\ 
%{\tiny \em   are potential}  
%\\
%{\tiny \em baryon-eaters }
%\\
%\rule{0mm}{1mm}
%}}} 

With the advent of the LED paradigm this situation dramatically changes.
First, the extended desert (14 orders of magnitude in energy!)
preceding the unification point of the gauge couplings
disappears. Indeed, in $4+n$ dimensions, at energies
$E\gsim 1/\delta$, the logarithmic law of running
 gives place to a power law.
(Here $\delta$ is the wall thickness which is usually assumed
to be related to $M_{\rm f}$.) If unification of the gauge couplings takes place
\cite{gher},
it occurs at energies close to  $M_{\rm f}$. With the fundamental scale 
$M_{\rm f}$ as low as
10 TeV in the ADD scenario, a disaster seems inevitable.
With new interactions  and  particles (mediating the proton decay)
as light as  10 TeV protons will decay immediately
unless special protective mechanisms
are found. 

The proton instability 
due to quantum gravity 
 is, at the very least, as severe a problem as the one discussed above.
If strong gravity occurs at the scale $M_{\rm f}\sim $
10 TeV, virtual black holes will be abundant
and they will destroy all globally conserved quantum numbers
(such as the baryon number) much faster than in $10^{32}$ years.
This is seen from  Eq.
(\ref{protltp}) where $M_{\rm Pl}$ is to be replaced by $M_{\rm f}$.
Needless to say,  this would be a mortal blow to the
ADD scenario.

\vspace{0.5mm}

One of possibilities to suppress the proton decay
 is to associate a gauge symmetry with the baryon number,
 more exactly, a discrete gauge symmetry\,.\footnote{What is a discrete gauge symmetry?
 One first gauges the baryon charge in a regular way coupling it, say, to a U(1) gauge boson.
 Then this continuous gauge symmetry must be spontaneously broken by a Higgs mechanism down
 to a discrete subgroup.} This mechanism of protection was invented long ago
 \cite{krwi}. As was noted by Krauss and Wilczek  \cite{krwi},
``neither black-hole evaporation, wormholes, nor anything else can violate discrete gauge symmetries."
With the advent of the LED paradigm people used it first
to guarantee the proton stability. This method 
was shown to be viable, i.e. its phenomenological implications are compatible
with experimental data \cite{pdrevp}. 

Here we will focus on an alternative, geometric protection,
one of many manifestations of a universal geometrical idea which, 
being combined with the ADD idea, brings lavish fruits.
It goes under the name {\em fat branes}, or branes with a substructure,
and was put forward by Arkani-Hamed and Schmaltz \cite{AH1999}
in the context of the problem of proton stability.\,\footnote{These authors also considered 
the hierarchy problem following the same
line of reasoning.}
%
%\marginpar{\frame{\shortstack{ 
%\rule{0mm}{1mm}\\
%{\tiny \em Domain walls\, } 
%\\ 
%{\tiny \em   may have a}  
%\\
%{\tiny \em substructure }
%\\
%\rule{0mm}{1mm}
%}}}

Let us examine the transverse structure
of the domain wall. A hypothetical slice through the wall is schematically
shown in Fig. 13. The overall wall thickness is $\delta$.
A crucial observation is that quarks and leptons 
need not be localized at one and the same point on the $Z$ axis,
and, moreover, the localization width need not coincide with 
$\delta$. Assume that quarks are localized on the left edge of the wall,
with the localization width $\ell \ll \delta$, while leptons are localized
on the right edge. If $W$ and other gauge bosons 
are smeared everywhere inside the wall
--- i.e. their degree of localization is $\delta$ --- this will
ensure that ``normal'' SM quark/lepton decays
proceed as they should. At the same time, the baryon number changing
transition of quarks into an appropriate lepton
will be suppressed by an exponentially small overlap of the wave functions
in the $Z$ direction. The suppression factor in the amplitude is
proportional to
\beq
\int \, d^n Z\, \Psi_q^3 (Z)\, \Psi_l^*(Z) \, \sim \,  e^{-\delta/\ell}\,,
\label{overint}
\eeq
where the subscripts $q$ and $l$ stand for quarks and leptons, respectively.
A relatively modest geometric hierarchy of $\delta/\ell \sim 30$,
after the exponentiation, will suppress the proton decay 
to the acceptable level, with no protection in the form of additional
(gauged) conservation laws. 
 
 \begin{figure}[h]
  \begin{center}
  \includegraphics[width=2.5in]{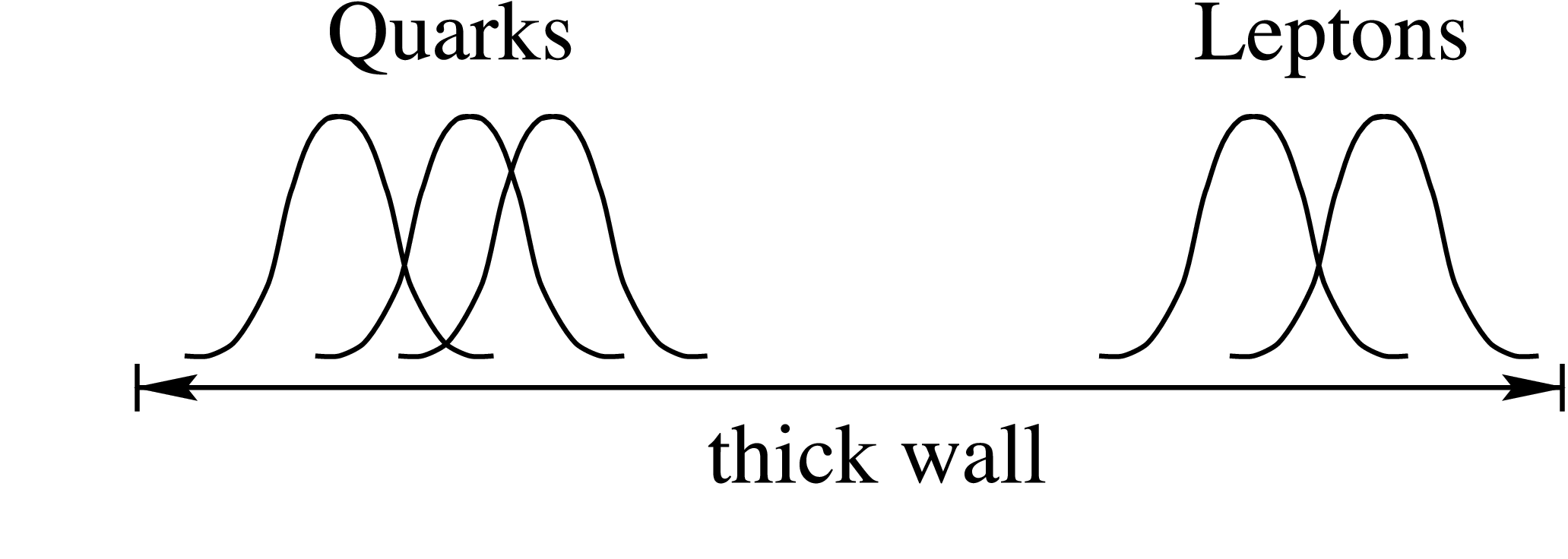}
\caption{A layered domain wall with a substructure.}
   \end{center}
\label{fat}
\end{figure}

A possible substructure of the domain wall defining our brane world
may provide geometric solutions to other problems from
flavordynamics, in particular those listed in Sect. \ref{phenim}.
For instance, the four-layer structure shown in Fig. 14
may explain \cite{DS2000} not only the fact of three distinct generations,
but also the pattern of quark masses and mixing angles \cite{DS2000,Matti}.
Four domains correspond to localizations
of the Higgs field and quark-lepton fields belonging
to generations three, two, and one.
It is clear that the suppression of mass terms due to 
overlap of the wave functions similar to Eq. (\ref{overint})
will depend on the number of the generation at hand ---
the closer it is to the Higgs brane, the larger is the overlap with the
Higgs field $Z$-profile,
implying a larger mass mass term. In this way a typical pattern of the
quark masses in three generations naturally emerges.

 \begin{figure}[h]
  \begin{center}
  \includegraphics[width=2.5in]{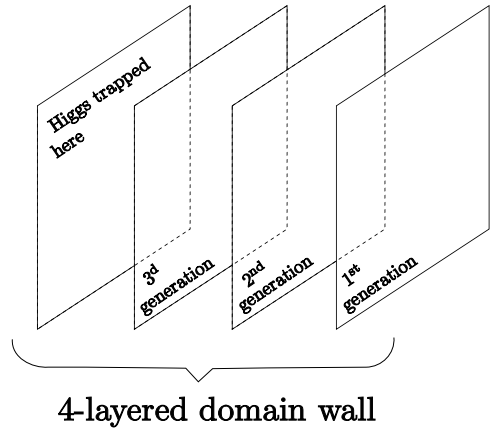}
\caption{Four-layer structure of the domain wall generates a reasonable pattern of the quark masses and mixing angles.}
   \end{center}
\label{fourlayer}
\end{figure}

\subsection{The lightness of the left-handed neutrino}
\label{llhn}

The masses of ``our'' left-handed neutrinos are believed to
lie in the ballpark of $10^{-2}$ or $10^{-3}$ eV.
How does the good old paradigm cope with
such small masses?

The standard explanation relies on the  seesaw  mechanism
invented in the late 1970's~\cite{mgm}.
The essence of the seesaw mechanism is as follows. Let us
assume that the Dirac
 neutrino mass is described by the term
$\mu\, \bar\nu_R\,\nu_L$ where the mass term $\mu$ is
of the order of $M_{\rm SM}$, its natural order of
magnitude. The right-handed neutrino which is (nearly) sterile
has a Majorana mass term $M\, \bar\nu_R\,\nu_R^c$
where the superscript $c$ stands for the charge conjugation.
It is natural to assume that $M\sim M_{\rm GUT}$.
Then, upon diagonalization of the mass matrix,
one finds that the true left-handed neutrino
is a mixture of $\nu_L$ and $\nu_R^c$
(the admixture of $\nu_R^c$ is very small, $\sim \mu /M$)
and its mass  is
\beq
m_{\nu_L} \, \sim \, \frac{\mu^2}{M} \,.
\label{lhnm}
\eeq
With the above assumptions regarding $\mu$ and $M$,
we get ``our'' neutrino mass in the right ballpark.
The lightness of ``our'' neutrino is due to the enormity of
$M_{\rm GUT}$.

As we remember, in the ADD scenario the {\em bona fide}
fundamental scale $M_{\rm f}$ is much lower. Gone with 
$M_{\rm GUT}$ is the seesaw mechanism. A
 question that immediately comes to one's mind is:
``Can one 
engineer a LED-based  mechanism (preferably, geometrical) that
would naturally explain the lightness of the left-handed neutrino?''

The answer to this question is positive, and, in fact, there exists
more than one solution. One of the nicest ideas belongs to the authors
of the ADD scenario themselves~\cite{addln}.
The right-handed neutrino carries no charges with respect to the SM gauge bosons.
Therefore, nothing precludes us from letting the right-handed neutrino escape
to the bulk, unlike the left-handed neutrino, which has to be localized
on the wall. 
The very existence of the wall may be responsible for the
disparity between 
$\nu_R$ and $\nu_L$. Indeed, the topology of the
wall solution (i.e. wall vs. anti-wall)
is typically correlated with the chirality of the fermion zero modes. Thus, 
it is quite natural to expect that a wall traps $\nu_L$, while 
$\nu_R$ is free to go in the bulk.
An anti-wall would trap $\nu_R$.

If the right-handed neutrino wave function is smeared all over the bulk
then the neutrino mass term, which is proportional to the
overlap of $\Psi_{\nu_L}$ and $\Psi_{\nu_R}^*$, 
takes the form
\beq
\int\,d^4 x\, H(x)\, \Psi_{\nu_L} (x)\, \Psi_{\nu_R}^* (x, Z=0)\,,
\label{ahsmt}
\eeq
where $H(x)$ is the Higgs field wave function. Due to the fact that
$\Psi_{\nu_R}$ is totally delocalized in the extra dimensions,
\beq
\Psi_{\nu_R} (x,Z)\,  \sim \,\frac{1}{\sqrt V_n}\, , 
\eeq
so that our neutrino mass gets a natural suppression
factor $V_n^{-1/2}$,
\beq
m_\nu \sim \frac{v }{\sqrt{V_n M_{\rm f}^n}}\, \sim\,  \frac{v\, M_{\rm f}}{M_{\rm Pl}}\,,
\eeq
where $v$ is the expectation value
of the Higgs field and we used Eq. (\ref{exdvol}). 
Given the uncertainty in numerical factors, this estimate places the neutrino mass
in the right ballpark.

Thus, the lightness of the left-handed neutrino is due to the
same reason why gravity is weak --- large volume of the extra dimensions. 

\subsection{Downside of the ADD scenario}
\label{dsadds}

Precision electroweak measurements firmly established the validity of the standard model.
What will
come beyond the standard model? Although theoretical speculations are abundant,
experimental support is scarce. In fact, the only semiquantitative achievement
in this direction  is the success of the gauge 
coupling unification.

The gauge interactions unified by the standard model are
SU(3)$\times$SU(2)$\times$ U(1). The first is the color group,
the last two represent electroweak interactions. At low energies the corresponding
three gauge couplings are very different in their values, see Fig. 9. The SU(2) and U(1)
couplings are measured with high precision;
the accuracy of the SU(3) coupling is not that high, mainly due to 
theoretical uncertainties in its determination. 
The logarithmic running brings them closer, and 
eventually all three intersect --- nearly
exactly at one point  ---  $M_{\rm GUT}$, which, in turn, turns out to be rather close to
$M_{\rm Pl}$. 
Figure 9 illustrates this statement.
It is important to note that the intersection occurs
only provided that SM is supersymmetrized. Thus, the above success may be viewed, 
simultaneously, as a semidirect indication that supersymmetry is relevant to nature.
It would be a pity to loose this encouraging indication. 

The unification and the desert paradigm
are closely connected.
Being honest, we should admit that the ADD scenario 
 erases the above success. The logarithmic running
stops at $M_{\rm f}$, where the three gauge couplings
are still very far from each other. Will a power-like running, 
which replaces the logarithmic one above $M_{\rm f}$, unify the couplings,
 and at what scale?
With the loss of the great desert
the answer to this question becomes model-dependent and almost completely devalued
\cite{gher}.

\section{A few words on other scenarios}

Next, we will outline, very briefly other scenarios, in which extra dimensions are instrumental in understanding the world we observe around us. Chronologically they appear later than ADD
and are ``ideologically" related.

\subsection{Warped Extra Dimensions}
\label{warped}

In the ADD scenario gravity of the domain wall as such plays little
role. Given that the brane tensions are small,
this is a good approximation. However, this need not be the case.
In the scenario suggested by Randal and Sundrum (RS) \cite{RS1,RS2}
the brane-induced 
 gravity strongly warps extra dimensions which  is instrumental in ensuring an
appropriate localization.

\begin{figure}
\begin{center}
\includegraphics[width=6cm]{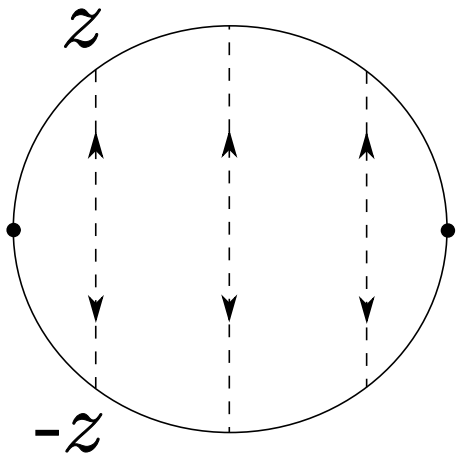}
\end{center}
\caption{Graphical representation of $S_1/Z_2$.
The dashed lines indicate the points of the circle which are to be identified.}
\label{orbi}
\end{figure}

A general observation on which the RS construction is based is as follows:
 if one has 3-branes in five dimensions, one can, in principle,  balance
the gravitational effects of the above  branes by a five-dimensional  bulk cosmological constant to get a theory in which an effective cosmological constant of our four-dimensional world 
 will vanish.  Our universe will seem
 static and flat for an observer on our brane~\cite{Rubakov:bb}. The price to pay for 
 this fine-tuning is a highly curved five-dimensional background.
 This phenomenon is called ``off-loading."
We off-load curvature from the brane on which we live onto the bulk.
Needless to say, there is no theoretical rationale for the above fine-tuning.
We just take the fact that the world we observe is (nearly) flat as given.

Let us discuss some basic elements of the RS construction.
One starts from a finite length  extra dimension~\cite{RS1} assumed to be an $S_1/Z_2$ orbifold.
What is this? Take a sphere $S_1$ and identify 
opposite points, as shown in Fig. 15. 
The points $0$ and $\pi$ are called fixed points since they are 
identified with themselves. Two branes are placed at these points in the extra dimension.
The solution of five-dimensional gravity -- I will not derive it here -- is self-consistent
(and consistent with the fine tuning discussed above) provided that one of the branes
has a positive tension while the tension of the other brane is negative
(and the same in the absolute value).

Since we want the Lorentz
invariance to be preserved in our four-dimensi\-onal world we have to assume
that
the induced metric (its $\mu\nu$ part)
at every point along the extra dimension is proportional to
the four-dimensional Minkowski metric $\eta_{\mu\nu}$, while the components of the five-dimensional 
 metric depend only on the 
fifth coordinate $Z$ in the following way: 
\begin{equation}
ds^2=e^{-A(Z)} dx^\mu dx^\nu \eta_{\mu\nu} -dZ^2\,.
\label{mm24}
\end{equation} 
The degree of warping along the extra dimension depends on the factor $e^{-A(Z)}$,
which is therefore called the warp factor. 
In the RS solution the warp factor depends on $Z$ exponentially,
\beq
A(Z) = 2\, k\, |Z|
\label{mm25}
\eeq
where $k$ is a constant of dimension of mass
which sets the scale of warping. It turns out that it is related to the five-dimensional
cosmological constant $\Lambda_{5D}$ and the five-dimensional
Planck constant $M_{5D}^{\rm Pl}$ by the formula \cite{RS1}
\beq
k^2 = -\frac{\Lambda_{5D}}{\left(M_{5D}^{\rm Pl}\right)^3}\,. 
\label{mm26}
\eeq
This relation implies, in particular, that
$\Lambda_{5D}$ is negative. Thus, the five-dimensional space-time one deals
with in the RS scenario is anti-de Sitter.

Now finally we are in a position to understand the impact of two branes -- a positive tension
brane at $Z=0$ and a negative tension brane at $Z=b$ --
as well as the impact of warping. Equations (\ref{mm24}) and  (\ref{mm25})
show that the induced metric on the negative tension brane
is exponentially smaller than that on the positive tension brane provided that $kb\gg1$,
\begin{equation}
g_{\mu\nu}^{\rm ind}\Big|_{Z=b}=e^{-2kb} \, \eta_{\mu\nu}\, .
\label{mm27}
\end{equation}
This suppression sets the scale for all other mass parameters. 
Indeed, consider the Higgs field part of the action. 
Proceeding to the canonically normalized kinetic term of the
Higgs field we see that the physical value of its vacuum condensate is 
``warped down'' to 
\begin{equation}
{v}_{\rm phys.Higgs}=e^{-kb} v_0\,,
\label{mm28}
\end{equation}
which shows, in turn, that all masses  following from
(\ref{mm28}) are exponentially suppressed on the negative tension brane (but not on the positive tension
brane). Then it is natural to refer to the  positive tension brane at $Z=0$ as the {\it Planck brane}, since the 
fundamental mass scale there is of the order of the Planck scale. The 
negative tension brane is said to be the {\it TeV brane} which follows from Eq. (\ref {mm28})
where we use $kb \sim 17$ and $M_{5D}^{\rm Pl} \sim M_{4D}^{\rm Pl}$.

In fact, there are
two popular versions of the RS scenario. The one discussed above (RS1), has a  finite size  extra dimension with two branes, one at each end. The second version (RS2) is similar to the first, but one brane has been placed infinitely far away, so that there is only one brane left in the model.
The particles of the standard model are placed on the Planck brane. This model was originally of interest because it represented an infinite five-dimensional model resembling, in many respects, 
 a four-dimensional model. 

For a more detailed consideration the reader is referred to numerous reviews,
for instance [35].

\subsection{Braneworlds with Infinite Volume Extra Dimensions}

Here we will briefly outline a model proposed by Dvali, Gabadadze, and Porrati (DGP)\,\cite{Dvali:2000hr} in 2000 
 in which, strictly speaking, gravity never becomes four-dimensional,
no matter how far in the infrared we go. 
Therefore, this is not a genuine compactification.
Unlike the ADD scenario in this construction
gravity only approximately imitates four-dimensional behavior at large distances. 
A residual ``tail" associated with extra dimensions produces an effect
which could, in principle,  reproduce the cosmic acceleration of dark energy \cite{Dvali:2002pe}.
Then the four-dimensional cosmological constant need not be nonvanishing.
The DGP mechanism  allows the volume of the extra space to be 
infinite,
\beq
V_N\equiv \int d^N\, Z \sqrt {G}\to \infty \,,
\label{infvolume}
\eeq
where $N$ is the  number of extra dimensions and $G$ is the metric tensor
in the $4+N$ dimensional space. For orientation below we will assume $N=1$,
although other versions of the DGP scenario, with $N>1$, were also considered.

The basic idea behind this scenario is as follows. We start from $4+N$-dimensional space,
with $4+N$-dimensional gravity governed by the standard Einstein--Hilbert action.
A four-dimensional domain wall is embedded in $N$ dimensions. 
As in the ADD scenario, all matter fields are assumed to be localized on the wall.
The central point of the DGP construction is
the emergence of the induced four-dimensional Einstein--Hilbert term on the wall
due to a loop of virtual matter localized on the wall, see Fig. 16. 
The bare action has no such term. However, as soon as the localized matter fields
are coupled to $4+N$ dimensional gravity, we obtain 4-dimensional gravity action on the wall 
from loops.
This mechanism described by Sakharov\,\cite{Sakharov} long ago 
is usually referred to as Sakharov's induced gravity. 

\begin{figure}
\begin{center}
\includegraphics[width=7cm]{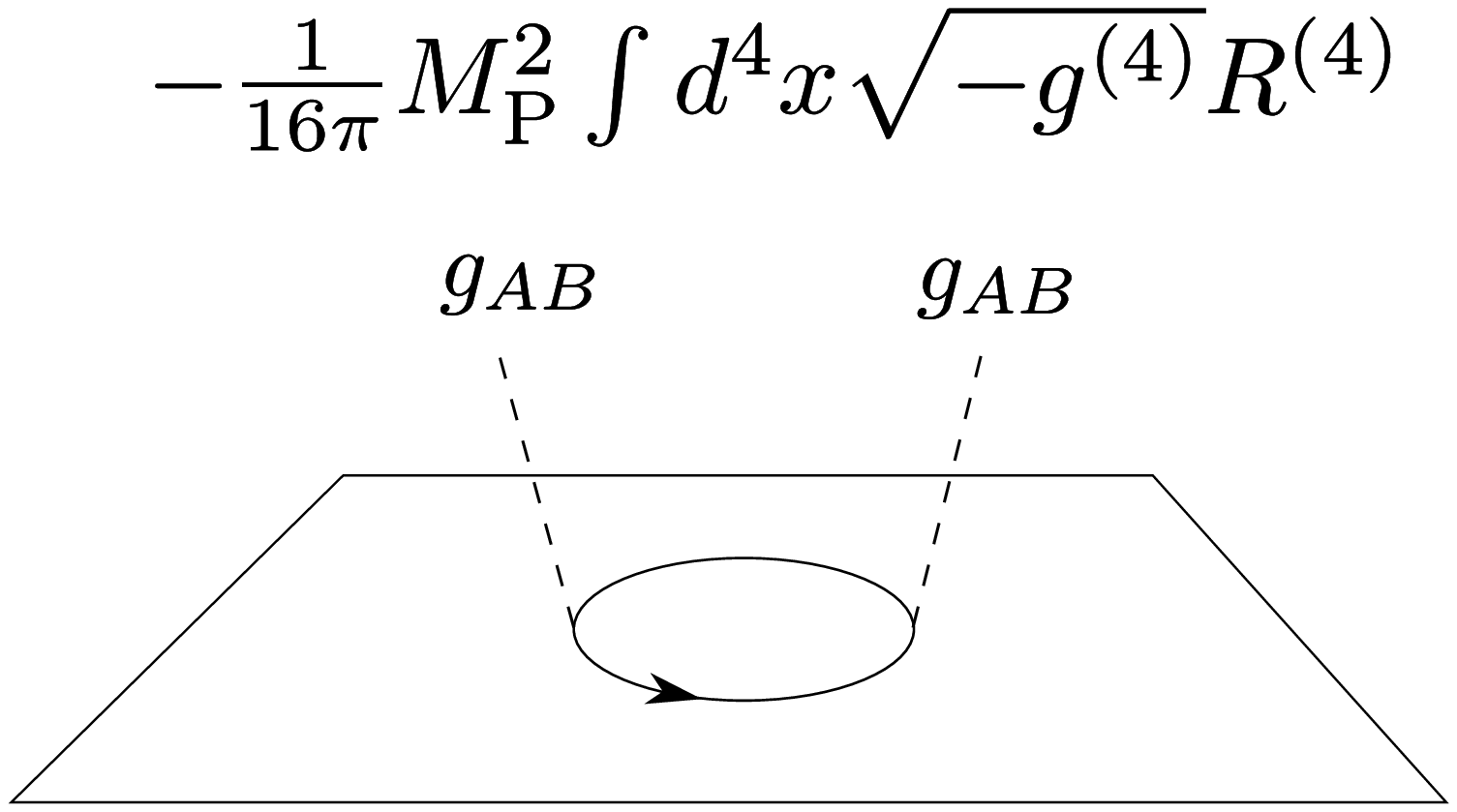}
\end{center}
\caption{One-loop contribution of
the matter fields to the effective action of gravitons.
The matter fields are localized on the brane yielding the four-dimensional
 Einstein--Hilbert term which is also localized on the brane.}
\label{dgpfi}
\end{figure}

Then the DGP action takes the form
\beqn
S
&=&
{\left(M_{5D}^{\rm Pl}\right)^3\over 2}\,\int d^4x \int_{-\infty}^{+\infty}dy
\sqrt{G}\, R_{5} 
\nonumber\\[4mm]
&+ &
\int d^4x \sqrt{g}\left[\frac{
\left(M_{4D}^{\rm Pl}\right)^2}{ 2}\, R_4+ 
\Lambda + {\cal L}_{\rm matter}(\Phi_{\rm SM},\Psi_{\rm SM})\right]\,,
\label{actDGP}
\eeqn
 where $G$ and $g$ stand for five- and four-dimensional metrics, respectively,
 $$
 g= G(Z=0)\,,$$
 $R_5$ and $R_4$ are the corresponding Ricci tensors, while
 $\Phi_{\rm SM}$, $\Psi_{\rm SM}$ is a generic notation for all boson and fermion matter fields.
 The parameter  $\left(M_{4D}^{\rm Pl}\right)^2$ is obtained  in Eq.~(\ref{actDGP})
 as an ultraviolet cut-off
 in loops of the brane-confined matter, representing the scale of energies at which
 the standard-model physics drastically changes. In the DGP scenario it is
 assumed that 
 $$M_{5D}^{\rm Pl}/M_{4D}^{\rm Pl}\gg 1\,.$$
 
Let us discuss the value of the domain wall tension.  One assumes that
the $4+N$-dimensional theory is supersymmetric, and that supersymmetry is
spontaneously broken only on the brane. 
Non-breaking of supersymmetry in the bulk is only 
possible due to the infinite volume of the extra space;
supersymmetry breaking is not transmitted
from the brane into the bulk since 
the breaking effects are suppressed by an infinite volume factor.
Then, the  bulk cosmological term can be set to zero, without any
fine-tuning. As for the  four-dimensional cosmological constant $\Lambda$
in Eq.~(\ref{actDGP}), the natural
value of  $\Lambda $ can be as low as TeV$^4$, since 
the brane tension can be protected above this value by 
${\mathcal N}=1$ supersymmetry. Please, note that
$\Lambda$ must be 
 fine-tuned in such a way that $\Lambda+\langle {\cal L}_{\rm matter} \rangle =0 $.
This is a usual fine-tuning of the cosmological constant. 
The DGP construction adds nothing new in this respect.

With the action (\ref{actDGP}) gravity from the sources localized on the domain wall
will propagate both, in the bulk and in the brane. Interplay of these two 
modes of propagation leads to quite a peculiar
 gravitational dynamics on the brane. 
Unlike the ADD scenario, in which gravity deviates from its four-dimensional form 
only at short distances,
in the DGP case deviations from the 
four-dimensional Newton law occur both at short and large distances.
Despite the fact that the volume
of extra space is infinite, an observer on the brane measures
four-dimensional gravitational interaction up to 
distances
\beq
r_c \sim \frac{
\left(M_{4D}^{\rm Pl}\right)^2
}{\left(M_{5D}^{\rm Pl}\right)^3}
\,.
\eeq
At distances larger than $r_c$ the Newton potential losses its four-dimensional form.
In order for the late-time cosmology to be standard 
we must require $r_c$ to be of the order of the universe size,  i.e.
$r_c ~\sim ~H_0^{-1}\sim 10^{28}~{\rm cm}$.
This, in turn, requires $M_{5D}^{\rm Pl}$ to be very small.
Small $M_{5D}^{\rm Pl}$ means that gravity in the
bulk  is strong. However, the large Einstein--Hilbert  term
on the brane ``shields'' matter localized on the brane
from strong bulk gravity.
 
\subsection{Universal extra dimensions}

For completeness I mention the universal extra dimension (UED) scenario \cite{ACDob}.
It assumes compactification of the ADD type. 
However, all standard matter fields are free to propagate through all of the extra-dimensional space
(which is essentially flat),
rather than being confined to a brane. The size of the extra dimensions is assumed to be in the ballpark
of TeV$^{-1}$. To my mind this scenario lacks elegant theoretical features
that can be found in ADD, RS and DGP. Phenomenology that ensues was discussed in the literature, see e.g. the review papers
[40].

\section{Conclusions}

When I googled {\em Large Extra Dimension Scenarios} I got $\sim 10^6$ hits. Needless to say, my
relatively short
introductory  lecture is way below the level that would allow you easy navigation in this ocean.
Hundreds of important papers  published after 1998
focus on this topic -- large extra dimensions -- from field theory, string/D-brane theory and phenomenology sides.
I did my best to acquaint you with the basic notions. Those of you who want to work in this direction should start reading the review papers mentioned above -- they contain representative lists
of the original literature. Others can just enjoy the idea, relax and wait for the
news (good or bad for the LED paradigm) which will hopefully come from LHC shortly.

\subsection*{Acknowledgments}

I would like to thank G. Gabadadze for numerous discussions
and comments on the maniscript.
I am grateful to Andrey Feldshtein and Maxim Konyushikhin
for providing me with figures used in this Lecture. 

\vspace{2mm}

 This work  is supported in part by DOE grant DE-FG02-94ER408. 

%\cite{uedrev}. 


\newpage
\begin{thebibliography}{99}

\bibitem{K1}
Th. Kaluza, Sitzungsber. Preuss. Akad. Wiss. Phys. Math. Klasse 966  (1921)
[Reprinted, with the English translation, in {\em Modern Kaluza--Klein Theories}, Eds.
T. Appelquist, A. Chodos, P.G.O. Freund
(Addison-Wesley, Menlo Park, 1987), p. 61]. Theodor Kaluza, of K\"{o}nigsberg University,
submitted his paper to Einstein in 1919.
Einstein apparently had serious doubts as to its merits --- he forwarded it
for publication in 
the above journal only in 1921.

\bibitem{K2}
In 1926 Oscar Klein who at that time was in
Copenhagen, and the Russian physicist
H. Mandel  independently
rediscovered Kaluza's theory  (O. Klein, Z.F. Physik, {\bf 37}, 895 (1926)
and Nature, {\bf 118}, 516 (1926);
H.~Mandel, Z.F. Physik, {\bf 39}, 136 (1926)).
Klein's papers are reprinted  in {\em Modern Kaluza--Klein Theories}, Eds.
T. Appelquist, A. Chodos, P.G.O. Freund
(Addison-Wesley, Menlo Park, 1987), p. 76 and 88.

\bibitem{pauli}
W. Pauli, Ann. d. Physik, {\bf 18},  305, 337 (1933).

\bibitem{EB}
A. Einstein and P.G. Bergmann, Ann. Math. {\bf 39},
810 (1931)
[Reprinted, with the English translation, in {\em Modern Kaluza--Klein Theories}, Eds.
T. Appelquist, A. Chodos, P.G.O. Freund
(Addison-Wesley, Menlo Park, 1987), p. 89].

\bibitem{wittenone}
E.~Witten,
%``Search For A Realistic Kaluza--Klein Theory,''
Nucl.\ Phys.\ B {\bf 186}, 412 (1981).
%%CITATION = NUPHA,B186,412;%%

\bibitem{JSJS}
 J.~Scherk and J.~H.~Schwarz,
  %``Dual Field Theory Of Quarks And Gluons,''
  Phys.\ Lett.\  B {\bf 57}, 463 (1975).
  %%CITATION = PHLTA,B57,463;%%

\bibitem{Candelas:en}
P.~Candelas, G.~T.~Horowitz, A.~Strominger and E.~Witten,
%``Vacuum Configurations For Superstrings,''
Nucl.\ Phys.\ B {\bf 258}, 46 (1985).
%%CITATION = NUPHA,B258,46;%%

\bibitem{AlFa}
A.~E.~Faraggi,
{\em Superstring phenomenology: A personal perspective,}
in {\sl Proc. 2nd International Conference on Physics Beyond The Standard Model,}
(June 1999, Tegernsee, Germany),
 ``{\sl Beyond the Desert 1999}", 
Eds.  H.V. Klapdor-Kleingrothaus and I.V. Krivosheina (Bristol, IOP, 2000),
p. 335-357 [hep-th/9910042].
%%CITATION = HEP-TH 9910042;%%

\bibitem{ibanez}
D.~Cremades, L.~E.~Iba\~{n}ez and F.~Marchesano,
%``Standard model at intersecting D5-branes: Lowering the string scale,''
Nucl.\ Phys.\ B {\bf 643}, 93 (2002)
[hep-th/0205074];
%%CITATION = HEP-TH 0205074;%%
{\em More about the standard model at intersecting branes,}
in {\sl Proc. Int. Conf. on Supersymmetry and unification of fundamental interactions, DESY, Hamburg 2002, }
Eds. P. Nath and P. Zerwas,
Vol. 1, p. 492  
[hep-ph/0212048];
%%CITATION = HEP-PH 0212048;%%
{\em Towards a theory of quark masses, mixings and CP-violation,}
hep-ph/0212064.
%%CITATION = HEP-PH 0212064;%%

\bibitem{jpol}
J.~Polchinski,
%``Dirichlet-Branes and Ramond-Ramond Charges,''
Phys.\ Rev.\ Lett.\  {\bf 75}, 4724 (1995)
[hep-th/9510017].
%%CITATION = HEP-TH 9510017;%%

\bibitem{Anton}
I.~Antoniadis,
  %``A Possible new dimension at a few TeV,''
  Phys.\ Lett.\  B {\bf 246}, 377 (1990).
  %%CITATION = PHLTA,B246,377;%%

\bibitem{horwi}
P.~Ho\v{r}ava and E.~Witten,
%``Eleven-Dimensional Supergravity on a Manifold with Boundary,''
Nucl.\ Phys.\ B {\bf 475}, 94 (1996)
[hep-th/9603142].
%%CITATION = HEP-TH 9603142;%%

\bibitem{Akama}
K.~Akama,
%``An Early Proposal Of 'Brane World',''
Lect.\ Notes Phys.\  {\bf 176}, 267 (1982)
[hep-th/0001113].
%%CITATION = HEP-TH 0001113;%%

\bibitem{Visser}
M.~Visser,
%``An Exotic Class Of Kaluza--Klein Models,''
Phys.\ Lett.\ B {\bf 159}, 22 (1985)
[hep-th/9910093].
%%CITATION = HEP-TH 9910093;%%

\bibitem{Rubakov:bb}
V.~A.~Rubakov and M.~E.~Shaposhnikov,
%``Do We Live Inside A Domain Wall?,''
Phys.\ Lett.\ B {\bf 125}, 136 (1983).
%%CITATION = PHLTA,B125,136;%%

\bibitem{JR}
R.~Jackiw and C.~Rebbi,
%``Solitons With Fermion Number 1/2,''
Phys.\ Rev.\ D {\bf 13}, 3398 (1976)
[Reprinted in {\em Solitons and Particles}, Eds.
C. Rebbi and G. Soliani (World Scientific, Singapore, 1984),
page 331.]
%%CITATION = PHRVA,D13,3398;%%

\bibitem{Dvali:1996xe}
G.~R.~Dvali and M.~A.~Shifman,
%``Domain walls in strongly coupled theories,''
Phys.\ Lett.\ B {\bf 396}, 64 (1997),
[E] \ B {\bf 407}, 452 (1997)
[hep-th/9612128].
%%CITATION = HEP-TH 9612128;%%

\bibitem{Dvali:1996bg}
G.~R.~Dvali and M.~A.~Shifman,
%``Dynamical compactification as a mechanism of 
%spontaneous supersymmetry  breaking,''
Nucl.\ Phys.\ B {\bf 504}, 127 (1997)
[hep-th/9611213].
%%CITATION = HEP-TH 9611213;%%

%\bibitem{BPS}
%E.~B.~Bogomolny,
%%``Stability Of Classical Solutions,''
%Yad.\ Fiz.\  {\bf 24}, 861 (1976)
%[Sov.\ J.\ Nucl.\ Phys.\  {\bf 24}, 449 (1976);
%reprinted in  {\em Solitons and Particles}, Eds.
%C. Rebbi and G. Soliani (World Scientific, Singapore, 1984),
%page 389];
%%%CITATION = SJNCA,24,449;%%
%M.~K.~Prasad and C.~M.~Sommerfield,
%%``An Exact Classical Solution For The 'T 
%%Hooft Monopole And The Julia-Zee Dyon,''
%Phys.\ Rev.\ Lett.\  {\bf 35}, 760 (1975)
%[Reprinted in {\em Solitons and Particles}, Eds.
%C. Rebbi and G. Soliani (World Scientific, Singapore, 1984),
%page 530].
%%%CITATION = PRLTA,35,760;%%

%\bibitem{GL}
%Y.~A.~Gol'fand and E.~P.~Likhtman,
%%``Extension Of The Algebra Of Poincare 
%%Group Generators And Violation Of P Invariance,''
%Pisma Zh.\ Eksp.\ Teor.\ Fiz.\  {\bf 13},  452 (1971)
%[JETP Lett.\  {\bf 13}, 323  (1971); reprinted in
%{\em Supersymmetry}, Ed. S. Ferrara (North-Holland/World Scientific,
%Amsterdam -- Singapore, 1987), Vol. 1, p.7].
%%%CITATION = JTPLA,13,323;%%

\bibitem{ADD1}
N.~Arkani-Hamed, S.~Dimopoulos and G.~R.~Dvali,
%``The hierarchy problem and new dimensions at a millimeter,''
Phys.\ Lett.\ B {\bf 429}, 263 (1998)
[hep-ph/9803315].
%%CITATION = HEP-PH 9803315;%%

\bibitem{ADD2}
I.~Antoniadis, N.~Arkani-Hamed, S.~Dimopoulos and G.~R.~Dvali,
%``New dimensions at a millimeter to a Fermi and superstrings at a TeV,''
Phys.\ Lett.\ B {\bf 436}, 257 (1998)
[hep-ph/9804398].
%%CITATION = HEP-PH 9804398;%%

\bibitem{Hoyle:2000cv}
C.~D.~Hoyle {\em et al.},
%``Sub-millimeter tests of the gravitational 
%inverse-square law: A search  for 'large' extra dimensions,''
Phys.\ Rev.\ Lett.\  {\bf 86}, 1418 (2001)
[hep-ph/0011014];
%%CITATION = HEP-PH 0011014;%%
J.~Chiaverini, S.~J.~Smullin, A.~A.~Geraci, D.~M.~Weld and A.~Kapitulnik,
%``New experimental constraints on non-Newtonian forces below 100-mu-m,''
Phys.\ Rev.\ Lett.\  {\bf 90}, 151101 (2003)
[hep-ph/0209325];
%%CITATION = HEP-PH 0209325;%%
E.~G.~Adelberger  [EOT-WASH Group Collaboration],
%``Sub-millimeter tests of the gravitational inverse square law,''
hep-ex/0202008; For reviews see
%%CITATION = HEP-EX 0202008;%%
G.~Landsberg,
%``Extra dimensions and more.,''
hep-ex/0105039;
%%CITATION = HEP-EX 0105039;%%
J.~C.~Long and J.~C.~Price,
%``Current short-range tests of the gravitational inverse square law,''
hep-ph/0303057.
%%CITATION = HEP-PH 0303057;%%

\bibitem{gher}
K.~R.~Dienes, E.~Dudas and T.~Gherghetta,
%``Extra spacetime dimensions and unification,''
Phys.\ Lett.\ B {\bf 436}, 55 (1998)
[hep-ph/9803466];
%%CITATION = HEP-PH 9803466;%%
%``Grand unification at intermediate mass 
%scales through extra dimensions,''
Nucl.\ Phys.\ B {\bf 537}, 47 (1999)
[hep-ph/9806292].
%%CITATION = HEP-PH 9806292;%%

\bibitem{pdrev}
S.~Wiesenfeldt,
{\em Proton decay in supersymmetric grand unified theories,}
DESY-THESIS-2004-009.
  %%CITATION = DESY-THESIS-2004-009;%%
  
  \bibitem{pdrevp}
  P.~Nath and P.~Fileviez P\'erez,
  %``Proton stability in grand unified theories, in strings, and in branes,''
  Phys.\ Rept.\  {\bf 441}, 191 (2007)
  [arXiv:hep-ph/0601023].
  %%CITATION = PRPLC,441,191;%%

\bibitem{blholep}
Y.~B.~Zeldovich,
%``Charge Asymmetry Of The Universe Due To 
%Black Hole Evaporation And Weak Interaction Asymmetry''
Pisma Zh.\ Eksp.\ Teor.\ Fiz.\  {\bf 24}, 29 (1976)
[JETP Lett.  {\bf 24}, 25  (1976)];
%%CITATION = ZFPRA,24,29;%%
%``A New Type Of Radioactive Decay: Gravitational 
%Annihilation Of Baryons,''
Phys.\ Lett.\ A {\bf 59}, 254 (1976).
%%CITATION = PHLTA,A59,254;%%
%R.~Kallosh, A.~D.~Linde, D.~A.~Linde and L.~Susskind,
%``Gravity and global symmetries,''
%Phys.\ Rev.\ D {\bf 52}, 912 (1995)
%[hep-th/9502069].

\bibitem{wormhole}
L.~F.~Abbott and M.~B.~Wise,
%``Wormholes And Global Symmetries,''
Nucl.\ Phys.\ B {\bf 325}, 687 (1989);
%%CITATION = NUPHA,B325,687;%%
S.~B.~Giddings and A.~Strominger,
%``String Wormholes,''
Phys.\ Lett.\ B {\bf 230}, 46 (1989);
%%CITATION = PHLTA,B230,46;%%
%``Baby Universes, Third Quantization And The Cosmological Constant,''
Nucl.\ Phys.\ B {\bf 321}, 481 (1989);
%%CITATION = NUPHA,B321,481;%%
S.~R.~Coleman and K.~M.~Lee,
%``Wormholes Made Without Massless Matter Fields,''
Nucl.\ Phys.\ B {\bf 329}, 387 (1990).
%%CITATION = NUPHA,B329,387;%%

\bibitem{krwi}
L.~M.~Krauss and F.~Wilczek,
  %``Discrete Gauge Symmetry in Continuum Theories,''
  Phys.\ Rev.\ Lett.\  {\bf 62}, 1221 (1989).
  %%CITATION = PRLTA,62,1221;%%

\bibitem{AH1999}
N.~Arkani-Hamed and M.~Schmaltz,
%``Hierarchies without symmetries from extra dimensions,''
Phys.\ Rev.\ D {\bf 61}, 033005 (2000)
[hep-ph/9903417].
%%CITATION = HEP-PH 9903417;%%

\bibitem{DS2000}
G.~R.~Dvali and M.~A.~Shifman,
%``Families as neighbors in extra dimension,''
Phys.\ Lett.\ B {\bf 475}, 295 (2000)
[hep-ph/0001072].
%%CITATION = HEP-PH 0001072;%%

\bibitem{Matti}
  C.~Matti,
  %``CKM pattern from localized generations in extra dimension,''
  Eur.\ Phys.\ J.\  C {\bf 48}, 251 (2006)
  [arXiv:hep-ph/0606158].
  %%CITATION = EPHJA,C48,251;%%

\bibitem{mgm}
P.~Minkowski,
  %``Mu $\to$ E Gamma At A Rate Of One Out Of 1-Billion Muon Decays?,''
  Phys.\ Lett.\  B {\bf 67}, 421 (1977);
  %%CITATION = PHLTA,B67,421;%%
M.~Gell-Mann, P.~Ramond and R.~Slansky,
{\em Complex Spinors And Unified Theories,}
in Proc. Stony Brook Supergravity Workshop
(September 1979)  {\sl Supergravity},
Eds. P. van Nieuwenhuizen and  D.Z. Freedman
(North Holland, Amsterdam, 1979), p. 315-321;
R.~N.~Mohapatra and G.~Senjanovic,
  %``Neutrino mass and spontaneous parity nonconservation,''
  Phys.\ Rev.\ Lett.\  {\bf 44}, 912 (1980);
  %%CITATION = PRLTA,44,912;%%
T.~Yanagida,
%``Horizontal Gauge Symmetry And Masses Of Neutrinos,''
Prog.\ Theor.\ Phys.\  {\bf 64}, 1103 (1980).
%%CITATION = PTPKA,64,1103;%%

\bibitem{addln}
N.~Arkani-Hamed, S.~Dimopoulos, G.~R.~Dvali and J.~March-Russell,
%``Neutrino masses from large extra dimensions,''
Phys.\ Rev.\ D {\bf 65}, 024032 (2002)
[hep-ph/9811448].
%%CITATION = HEP-PH 9811448;%%

\bibitem{RS1}
L.~Randall and R.~Sundrum,
%``A large mass hierarchy from a small extra dimension,''
Phys.\ Rev.\ Lett.\  {\bf 83}, 3370 (1999)
[hep-ph/9905221].
%%CITATION = HEP-PH 9905221;%%

\bibitem{RS2}
L.~Randall and R.~Sundrum,
%``An alternative to compactification,''
Phys.\ Rev.\ Lett.\  {\bf 83}, 4690 (1999)
[hep-th/9906064].
%%CITATION = HEP-TH 9906064;%%

\bibitem{CCcc}
C.~Csaki,
{\em TASI lectures on extra dimensions and branes,}
  arXiv:hep-ph/0404096,
  %%CITATION = HEP-PH/0404096;%%
in {\sl Particle Physics and Cosmology}, Eds. H. Haber and A. Nelson
(World Scientific, Singapore, 2004), p. 605.

\bibitem{Dvali:2000hr}
  G.~R.~Dvali, G.~Gabadadze and M.~Porrati,
  %``4D gravity on a brane in 5D Minkowski space,''
  Phys.\ Lett.\  B {\bf 485}, 208 (2000)
  [arXiv:hep-th/0005016].
  %%CITATION = PHLTA,B485,208;%%
  
\bibitem{Dvali:2002pe}
 C.~Deffayet, G.~R.~Dvali and G.~Gabadadze,
  %``Accelerated universe from gravity leaking to extra dimensions,''
  Phys.\ Rev.\  D {\bf 65}, 044023 (2002)
  [arXiv:astro-ph/0105068];
  %%CITATION = PHRVA,D65,044023;%%
  G.~Dvali, G.~Gabadadze and M.~Shifman,
  %``Diluting cosmological constant in infinite volume extra dimensions,''
  Phys.\ Rev.\  D {\bf 67}, 044020 (2003)
  [arXiv:hep-th/0202174].
  %%CITATION = PHRVA,D67,044020;%%
  
\bibitem{Sakharov}
  A.~D.~Sakharov,
  %``Vacuum quantum fluctuations in curved space and the theory of
  %gravitation,''
  Sov.\ Phys.\ Dokl.\  {\bf 12}, 1040 (1968) [reprinted in  Sov. \ Phys. \ Usp. {\bf 34},  394 (1991)].
  %%CITATION = GRGVA,32,365;%%
  
  \bibitem{ACDob}
  T.~Appelquist, H.~C.~Cheng and B.~A.~Dobrescu,
  %``Bounds on universal extra dimensions,''
  Phys.\ Rev.\  D {\bf 64}, 035002 (2001)
  [arXiv:hep-ph/0012100].
  %%CITATION = PHRVA,D64,035002;%%
  
  \bibitem{uedrev}
    G.~D.~Kribs,
{\em Phenomenology of extra dimensions,}
  arXiv:hep-ph/0605325,
  in {\sl Physics in $D\geq 4$}, Ed. J. Terning, C. Wagner and D. Zeppenfeld,
  (World Scientific, Singapore, 2006), p. 633;
  %%CITATION = HEP-PH/0605325;%%
  D.~Hooper and S.~Profumo,
  %``Dark matter and collider phenomenology of universal extra dimensions,''
  Phys.\ Rept.\  {\bf 453}, 29 (2007)
  [arXiv:hep-ph/0701197].
  %%CITATION = PRPLC,453,29;%%

%\end{thereferences} 
\end{thebibliography}
\end{document}